*Author Names & Affiliations:*
D. Sousa[1*] and C. Small[1]
[1]Lamont-Doherty Earth Observatory, Columbia University, Palisades, NY 10964 USA
Email: d.sousa@columbia.edu. Tel.: +1 5303044992

*Abstract:*
        Many biogeophysical processes continuously evolve in both space and time. Some of these processes change gradually enough across both space and time to be oversampled by the satellite image archive. In these cases, image time series of Earth observations can support robust spatiotemporal analysis. In this analysis, we leverage the spectral differences of green vegetation, soil, and clouds to perform an EOF analysis which simultaneously characterizes monsoon cloud cover and vegetation phenology in a temperate cloud forest in Dhofar, southern Oman. The Normalized Difference Vegetation Index of daily MODIS reflectance clearly distinguishes clouds and bare soil from both grassland and cloud forest vegetation. Using the temporal feature space of the low order principal components of the NDVI time series, we identify distinct annual cycles of cloud cover and vegetation phenology corresponding to geographically contiguous areas of coupled monsoon-phenological activity (*phenoregions*). The EOF analysis characterizes the temporal dynamics of these phenoregions. The wadi cloud forests of the Jabal Al Qara and Jabal Al Qamar can each be described with 2 distinct spatiotemporal patterns of monsoon phenology, corresponding to the eastern and western portions of each range. In addition, the rangetop grasslands of the Jabal Al Qara show 3 distinct spatiotemporal patterns, corresponding to eastern, central, and western portions of the range. While each phenoregion has a clearly repeating annual cycle, each also demonstrates substantial interannual variability in both cloud cover and phenology. This interannual variability is of a sufficient magnitude to mask any trends in cloud cover or phenology in any of the regions in the Dhofar Mountains. Finally, using the degree of asymmetry of the cumulative NDVI time series, we infer a causal link between duration of cloud cover and amplitude of phenology.

*Introduction:*

Optical reflectance is a biogeophysically relevant quantity observed at increasing frequency in both space and time. The reflectance of the Earth surface is the fraction of incident sunlight at a given wavelength which is reflected towards an imaging sensor. Reflectance spectra of many Earth materials have been well characterized, allowing inference of the relative abundance of these materials from multispectral imagery.

The synoptic multispectral image archive provides decades of intercalibrated reflectance measurements. New satellite constellations add to this archive with measurements collected at increasing sampling frequency in both space and time. Some climatic and ecological phenomena vary gradually enough in both space and time to be effectively oversampled in the satellite image archive. For these processes, image time series of Earth observations – now freely available – are capable of supporting robust spatiotemporal analysis.

In this work, we leverage the known spectral differences of green vegetation, soil, and clouds to perform an EOF analysis which simultaneously characterizes monsoon cloud cover and vegetation phenology in a temperate cloud forest in Dhofar, southern Oman. We use the EOF analysis to characterize regions with statistically distinct monsoonal and phenological cycles and identify both gradations and abrupt transitions among these regions. We then quantify seasonal and interannual differences in monsoon timing and vegetation phenology within and among these regions. The simultaneous spatiotemporal characterization of cloud cover and vegetation abundance provides insight into both monsoon variability and the response of different vegetation communities to this variability. While we observe no significant interannual trends in either vegetation abundance or monsoonal cloud cover in any of these regions, we find consistencies between interannual variations of the two variables. We suggest that this method is sufficiently general to apply in other landscapes characterized by coupling between vegetation phenology and seasonal precipitation.

*Study area:*

**Figure 1. Index map showing Dhofar topography with khareef and cloud forest extents. Post-khareef vegetation (green) coincides with persistent khareef cloud cover (cyan/white).**

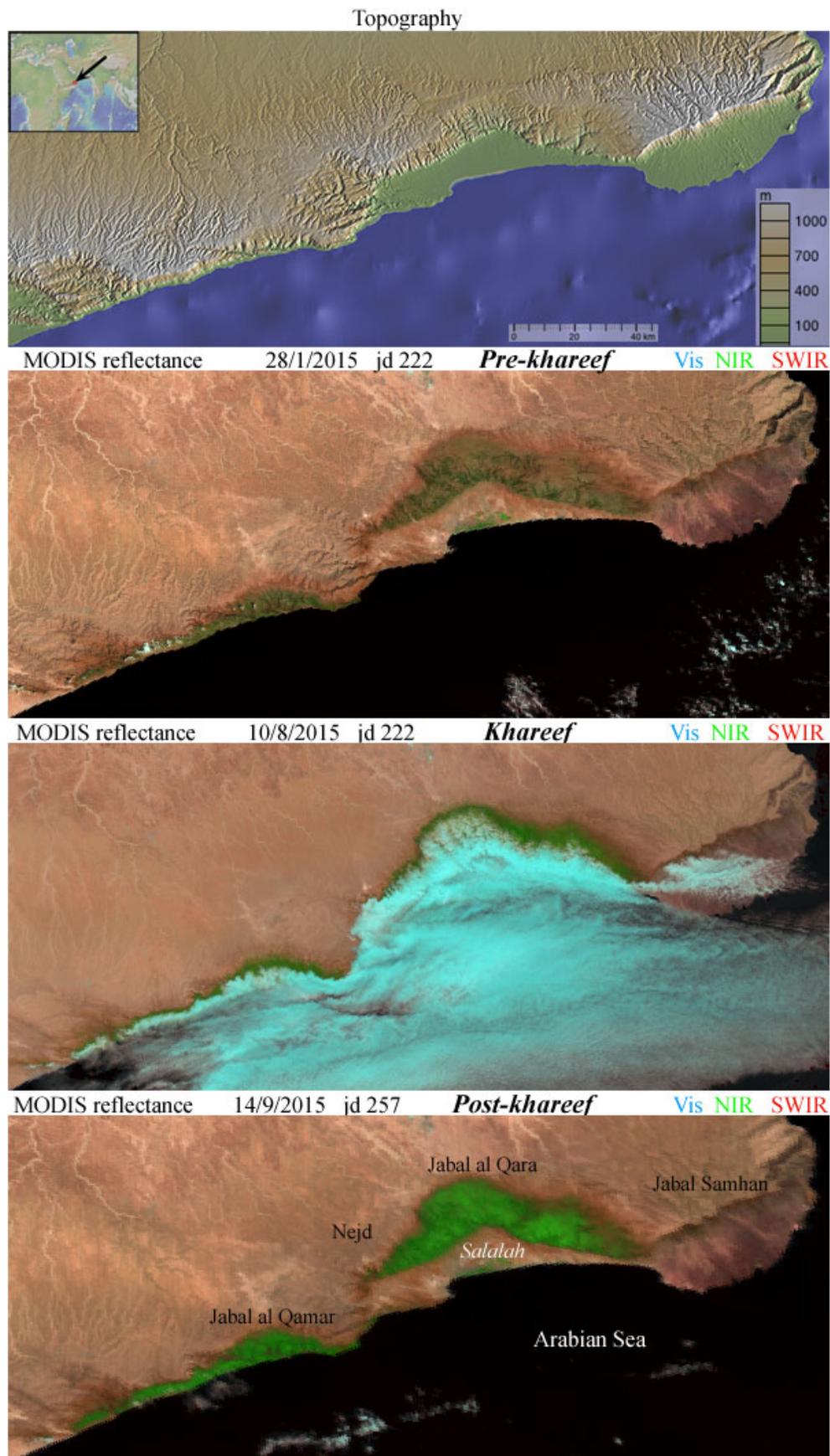

Figure 1  Index map showing Dhofar topography with khareef and cloud forest extents. Post-khareef vegetation (green) coincides with persistent khareef cloud cover (cyan/white).

The Dhofar mountains (*Jabal Dhofar*) of southern Oman are comprised of three ranges: the Jabal Al Qamar, Jabal Al Qara, and Jabal Samhan. The three ranges each rise abruptly from near sea level to elevations over 1000 m. The eastern Jabal Samhan is the highest of the three ranges, forming a steep forested slope which is home to a preserve for the Arabian Leopard [*Spalton and Al Hikmani*, 2014]. The Jabal Al Qara lies immediately west of the Samhan. The coastward face of the Qara is characterized by deeply incised valleys (*wadis*), while the rangetop forms a gradual ramp up into the persistently arid desert of the Nejd and the Empty Quarter (*Rub' al Khali*) of the Arabian Peninsula. Between the Jabal Al Qara and the Arabian Sea lies the Salalah Coastal plain. The western Jabal Al Qamar rises more abruptly from the sea and grades more rapidly into the Nejd. For an excellent introduction to the geography of the Jabal Dhofar, see [*Reade*, 1980]; for more general context see [*Thesiger*, 1948; 1949; *Thomas*, 1931].

The climate of Dhofar is dominated by the annual Arabian monsoon (*khareef*). The khareef is marked by persistent cloud cover from roughly mid-June to mid-September. The khareef brings nearly all of the water available to biota in the Jabal Dhofar. While some of this water is delivered in the form of rainfall, it has long been recognized that a substantial fraction of the moisture available to plants in this area condenses directly from the ground level clouds onto the plant and soil surfaces. This phenomenon is deemed *horizontal precipitation* and has been the subject of study for its importance for both the local ecohydrology [*Hildebrandt*, 2005] and its potential capture for anthropogenic uses (e.g. [*Abdul-Wahab et al.*, 2010; *Fallon*, 1978]).

The relative geographical isolation and distinct microclimate of the Jabal Dhofar contribute to high endemic biodiversity and a unique floral assemblage, documented in detail in [*Miller and Morris*, 1988]. The Jabal Samhan has only a narrow strip of monsoon cloud forest and so will be excluded in the remainder of this analysis. The coastward slopes of the Jabal Al Qara and Qamar are each home to a temperate cloud forest ecosystem dominated by the *Anogeisis dhofarica* tree. The rangetops of the Jabal Al Qara are characterized by broad, rolling grasslands which are grazed intensively by camels, cattle, feral asses, and goats.

The difference in species composition between the wadis and rangetops has been well documented by previous work. However, to our knowledge, neither the dynamics of the coupling between precipitation and phenology, nor the spatial variability within ecozones has yet been characterized in publication.

*Objectives:*

We simultaneously characterize the dominant spatiotemporal patterns of annual cloud cover and vegetation phenology in the temperate cloud forest of Dhofar by:
- Mapping distinct spatial patterns of coupled monsoon-phenological activity.
- Characterizing the average annual cycle and interannual variability of monsoon cloud cover and post-monsoon vegetation growth and senescence.
- Identifying trends, if any, in monsoon phenology over the 17 year duration of daily satellite observations.

We perform this characterization on a 17 year near-daily image time series of a single spectral index using the spatiotemporal analysis and Time-Space characterization methods described by [*Small*, 2012].

We find:

- The wadi cloud forests of the Jabal Al Qara and Jabal Al Qamar can each be described with 2 clearly distinct spatiotemporal patterns of monsoon phenology, corresponding to the eastern and western portions of each range.

- The rangetop grasslands of the Jabal Al Qara show 3 distinct spatiotemporal patterns, corresponding to eastern, central, and western portions of the range, while the rangetop grasslands of the Jabal Al Qamar are so compressed that they rapidly grade into the Nejd desert and are phenologically less distinct than those of the Qara.

- The spatiotemporal patterns which emerge from the EOF analysis correspond to spatially contiguous geographical regions which are phenologically distinct. We refer to these as *phenoregions*.

- While each phenoregion has a clearly repeating annual cycle, each also demonstrates substantial interannual variability in both cloud cover duration and phenologic response. We find this interannual variability to be of a sufficient magnitude to mask any obvious trends in monsoon cloud cover or vegetation phenology in any of the regions in the Dhofar Mountains.

- In the Jabal Al Qara, we also find that phenoregions with higher amplitude vegetation phenology tend to have longer duration of monsoon cloud cover. While this suggests a relationship between monsoon duration and vegetation abundance, the link between the two – quantity of water precipitated onto the land surface during the monsoon and stored in the soil after the monsoon – remains largely unknown.

***Data:***

    This study uses 17 years of near-daily surface reflectance imagery from the Moderate Resolution Imaging Spectroradiometer (MODIS) collection #6, acquired free of charge from the NASA REVERB data hub (https://reverb.echo.nasa.gov/). Data drop outs were excluded from this analysis by flagging acquisitions with only partial coverage of the study area (swath edges) and those with large areas of missing data, yielding a total of 4236 images. Data corruption, indicated by anomalous reflectance values, was particularly obvious in three years (2002, 2011, and 2016), as seen in several figures in this paper. These years were excluded from the calculation of the covariance matrices on which the EOF analysis was based, but they are included in the subsequent analysis because of the information provided by the non-corrupt acquisitions within each year.

    Field reflectance spectra of rock and soil substrates, green and senescent vegetation and clouds were collected using an ASD HandHeld2 field spectrometer. These Visible-Near Infrared (VNIR) field spectra were used for comparison of sub-meter scale reflectance of specific land covers with the aggregated reflectance measurements derived from the MODIS radiance observations. The details of this comparison are given by [Small et al (2017)]. Land cover and vegetation type were field validated by all 4 authors with geotagged photographs and dashcam video along an extensive set of driving and walking transects throughout the study area in February and March of 2016 and 2017. These *in situ* observations were used to scale field reflectance measurements to meter-resolution (Quickbird, Worldview) and simultaneously acquired decameter-resolution (Landsat 8, Sentinel-2) satellite imagery – also described in detail by [Small et al (2017)].

*Analysis:*

*Broadband Reflectance Spectroscopy*

**Figure 2. Reflectance of soil (S), cloud (C), green (GV) and non-photosynthetic (NPV) vegetation in the Dhofar study area. When viewed by a broadband sensor, like MODIS or Landsat, the true hyperspectral reflectance spectrum (thin) of a material is integrated with the sensitivity of each channel of the sensor, given by its spectral response (dashed). The result is a more discrete broadband spectrum (circles). The NDVI spectral index is a measure of the spectral slope between the integrated visible red and NIR wavelengths. Green vegetation has a large positive slope, but soils and NPV generally have much smaller positive slopes. Clouds are highly variable but generally show very small slopes near 0. The MODIS sensor images at 500 m resolution at SWIR wavelengths, but only the visible red and NIR are imaged at 250 m resolution.**

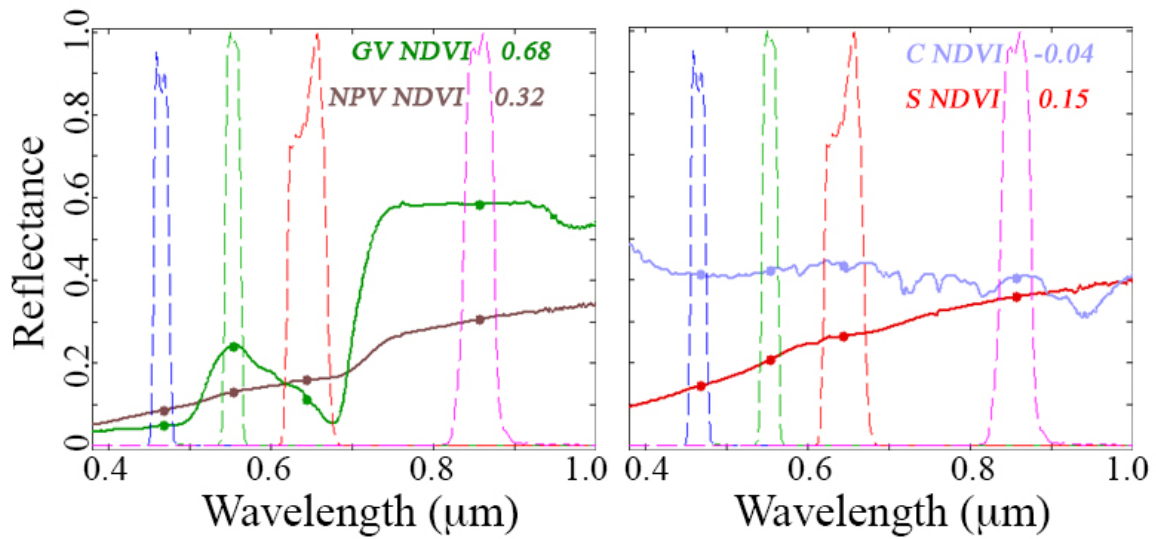

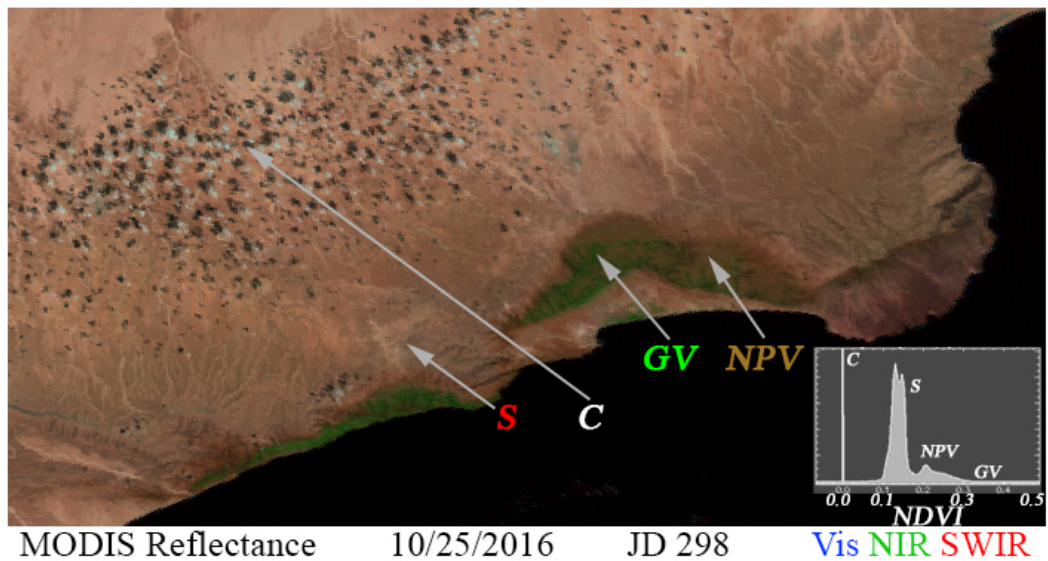

Figure 2. Reflectance of soil (S), cloud (C), green (GV) and non-photosynthetic (NPV) vegetation in the Dhofar study area. When viewed by a broadband sensor like MODIS or Landsat, the true hyperspectral reflectance spectrum (thin) of a material is integrated with the sensitiviy of each channel of the sensor, given by its spectral response (dashed). The result is a more discrete broadband spectrum (circles). NDVI is a normalized measure of the spectral slope between the integrated visible red and NIR wavelengths. Green vegetation shows a large positive spectral slope, but soils and non-photosynthetic vegetation generally also show positive spectral slopes. Clouds are highly variable but generally show smaller spectral slopes near zero. The MODIS sensor images at 500 m resolution at SWIR wavelengths, but only the visible red and NIR are imaged at 250 m.

The reflectance spectrum of a material is given by the fraction of incident light reflected off its surface into the aperture of a sensor over a range of wavelength bands. Example field spectra of four common land covers (including clouds) in Dhofar were collected during the 2017 field season and are shown as continuous curves in Figure 2. Photosynthetic vegetation appears green to our eyes because it is brighter in the visible green (~0.55 μm) than the visible red or blue. However, it is far brighter in the Near Infrared (NIR, 0.75 - 1.00 μm). Non-photosynthetic vegetation (NPV, e.g. senescent grass & leaf litter) and soil often have similar reflectance spectra, increasing in brightness (nearly) monotonically over the visible-NIR (VNIR) range of wavelengths. Clouds can have a wide range of reflectance spectra, dominated by the chemical absorptions of liquid and solid water particles of various size fractions. For an excellent introduction to the physics of reflectance spectroscopy and the optical properties of various materials see [*Hapke*, 2012]. For a comprehensive library of reflectance spectra of common earth materials see [*Kokaly et al.*, 2017].

The MODIS sensor is a broadband instrument which integrates the true continuous reflectance spectra of the materials imaged into a relatively small number (36) of bands. The response of the MODIS instrument over its four VNIR wavelength intervals (dashed lines) is also shown in Figure 2. The simulated broadband spectrum of the field spectra is represented by four dots. While substantially more information is contained in the continuous spectra, gross discrimination of major absorptions can often be achieved from broadband imagery – though with considerable nonuniqueness. Note, for instance, that minor absorptions distinguishing the continuous NPV and soil spectra descend into ambiguity when integrated by a broadband sensor.

Of the four MODIS VNIR bands, the red and NIR are imaged at (nominally) 250 m resolution, while the blue and green are imaged at 500 m resolution. While it is possible to combine the data streams of these bandpasses, and merged products are available, the result is inevitable blurring of spatial detail. In many applications this is acceptable (or even necessary). However, the spatial scale of the vegetation communities of the Jabal Al Qara and Jabal Al Qamar is sufficiently fine that any degradation of spatial resolution would result in substantial information loss. For this reason, we confine the analysis in this study to only the red and NIR bands collected at 250 m. We use a single spectral index to distinguish between rock and soil substrates, vegetation, clouds and water in each daily MODIS image. We use a continuous spectral index rather than a discrete land cover classification because we seek to use the information contained in the continuous variations of the index as a proxy for continuous changes in the land surface and cloud cover.

Spectral indices derived from only two broadband channels are inherently limited. By far the most common spectral index using red and NIR wavelengths leverages the normalized difference of the two channels to quantify the slope of the reflectance spectrum. The index most commonly used for this purpose is the Normalized Difference Vegetation Index (NDVI, [*Rouse et al.*, 1974]:

$$NDVI = \frac{NIR - Red}{NIR + Red}$$

NDVI has been shown to exhibit substantial nonlinearity as a metric for subpixel vegetation area (e.g. [*Small and Milesi*, 2013a; *Sousa and Small*, 2017]). For this reason, in this analysis we use NDVI only as a metric to quantify temporal change of the relative – not absolute – green vegetation abundance. Vegetation fraction as estimated by spectral mixture analysis has been shown to scale linearly with vegetated area [*Elmore et al.*, 2000; *C. Small*, 2001; *C Small and Milesi*, 2013b], without the nonlinear response to soil reflectance and atmospheric effects that complicate the use of NDVI (e.g. [*Myneni et al.*, 1995], [*Elvidge and Chen*, 1995]). Vegetation fraction is therefore a more robust and informative metric for vegetation abundance. However, spectral mixture analysis requires more than two wavebands and the coarsening of spatial resolution required would lose important detail in this study location.

Furthermore, examination of Figure 2 shows that NDVI also yields another important benefit – straightforward discrimination between presence and absence of cloud cover. Low level clouds are usually spectrally flat in the red and NIR, yielding index values near zero.  NDVI distributions in the study area are strongly multimodal with distinct modes separating clouds and substrates (rocks, soils and NPV).  The continuum of vegetation abundance is manifest as a long upper tail of higher NDVI values. This observation, combined with the spatiotemporal consistency of the monsoon, allows a single simple metric to capture features of both elements of the coupled monsoon-phenological system of the Jabal Dhofar cloud forest.

*Spatial patterns of temporal variability*
    **Figure 3. Spatiotemporal vegetation from MODIS NDVI time series.  A) Cloud forests generally have higher annual mean NDVI than grasslands because they senesce more slowly throughout the year. B).Time-Space cube of a subset of the Jabal Qara for 2006. Time series for pixels along the north and east edges of the cube face are shown on the sides. Plants senesce until the Khareef begins in June, then are obscured by cloud for ~2 months. Plants are maximally green when clouds clear after the khareef, then senesce at variable rates throughout the year. Rangetop grasslands senesce faster than wadi cloud forests. Daily NDVI time series for two sample locations (blue and red dots) showing daily observations with intermittent cloud cover (NDVI ~0) throughout the year.**

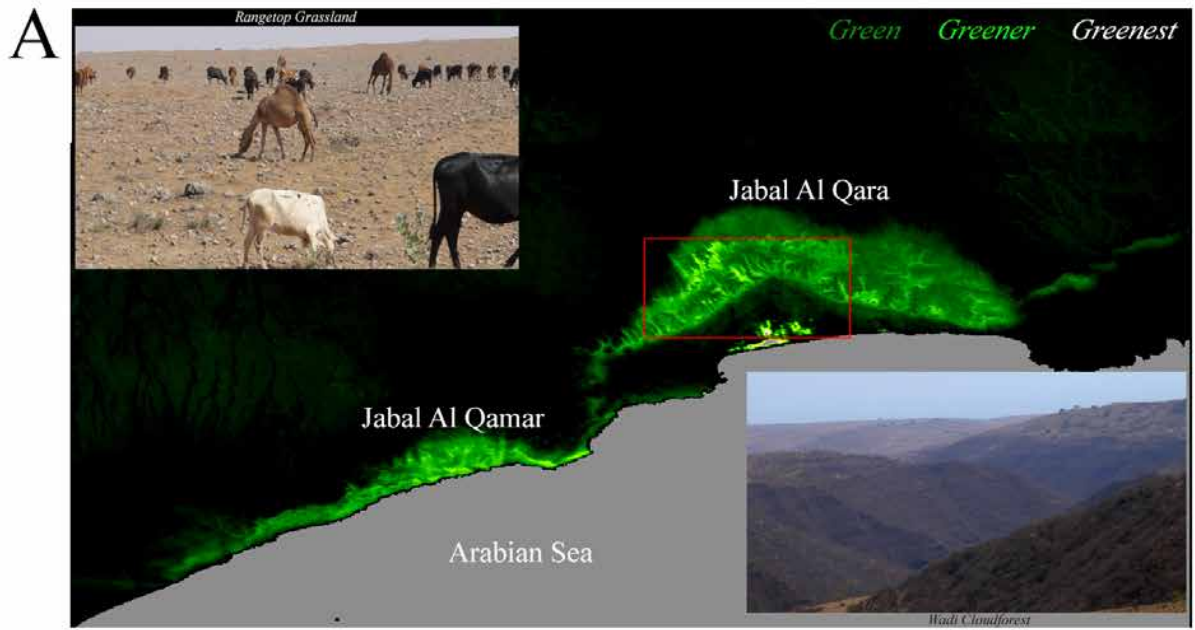

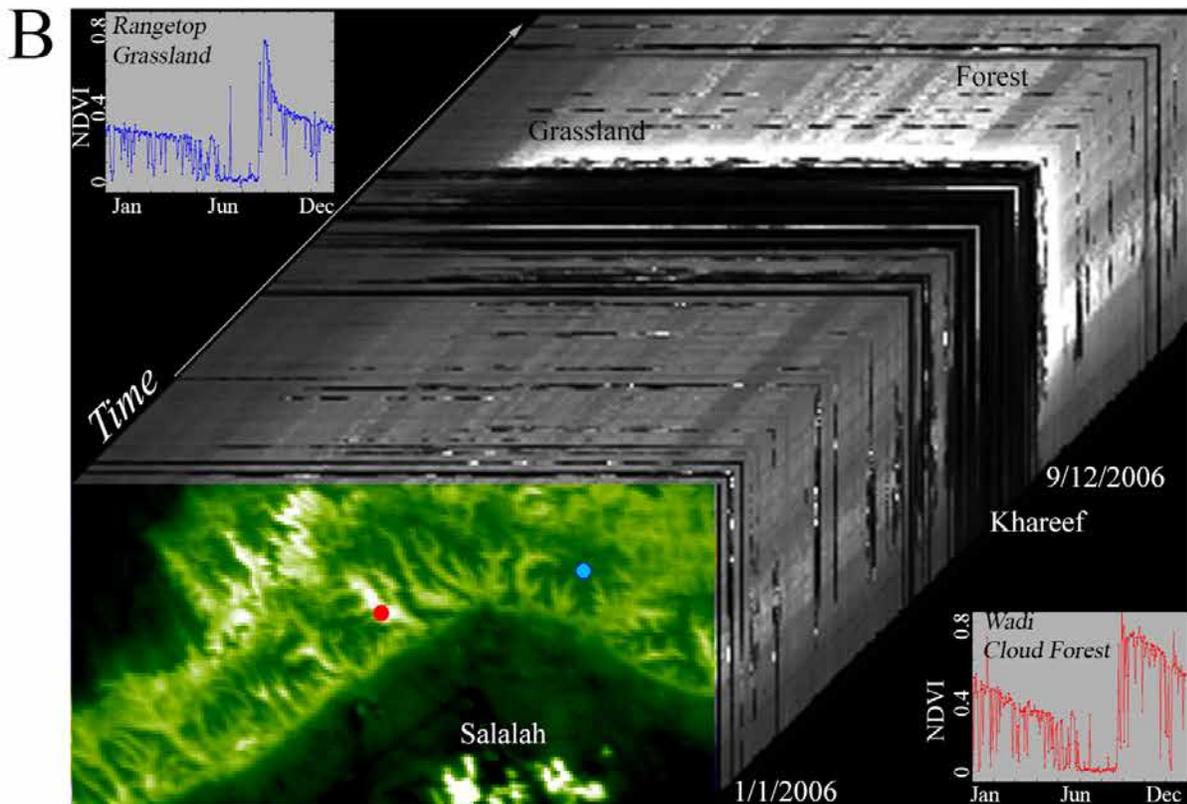

Figure 3. Spatiotemporal vegetation from MODIS NDVI time series.  A) Cloud forests generally have higher annual mean NDVI than grasslands because they senesce more slowly throughout the year. B).Time-Space cube of a subset of the Jabal Qara for 2006. Time series for pixels along the north and east edges of the cube face are shown on the sides. Plants senesce until the Khareef begins in June, then are obscured by cloud for ~2 months. Plants are maximally green when clouds clear after the khareef, then senesce at variable rates throughout the year. Rangetop grasslands senesce faster than wadi cloud forests. Daily NDVI time series for two sample locations (blue and red dots) showing daily observations with intermittent cloud cover (NDVI ~0) throughout the year.

The temporal mean of 17 years of near-daily NDVI observations compresses the overall spatial pattern of vegetation abundance in the Jabal Al Qamar and Jabal Al Qara into one image (Figure 3). In the Jabal Al Qara, the coastward-facing wadis appear as distinct, spatially contiguous areas of consistently high NDVI (bright), while the rangetops have substantially lower average NDVI because they senesce rapidly after the khareef. Field photos demonstrating the difference between the two regions show a marked difference in flora. The wadis are characterized by the *Anogeisis* dominated cloud forest ecozone, while the rangetops are seasonal grasslands subject to heavy grazing by macrofauna.

Example single year, single pixel time series for each ecozone are also shown in Figure 3. The rangetop grasslands show high NDVI immediately following the khareef, followed by rapid decrease as the grasses dry, senesce, and are grazed. The wadi cloud forests, in contrast, show even higher NDVI immediately following the khareef with a much slower, linear decay over the course of the year. In both ecozones, the khareef is distinctly visible as a period of consistently low NDVI.

The Time-Space cube shown in Figure 3 shows the spatial and temporal richness of the dataset. In this representation, the time series for a single example year are shown for pixels on the top and right edge of the face of the cube. The annual cycle is clearly present: starting at relatively low NDVI early in the calendar year, to near zero NDVI characteristic of cloud cover during the khareef, to higher then decaying NDVI immediately after the khareef. Occasional breaks in cloud cover during the khareef show that the vegetation greens while under cloud. In this year, the khareef has an abrupt end everywhere in the Jabal Al Qara. However, the onset is less clear, with sporadic cloudy days interspersed with clear days. Note also the near complete cloud cover during the khareef and spatial variability of cloud cover the rest of the year.

*Temporal patterns of spatial variability*

**\*\* Figure 4. Interannual comparison of annual NDVI time series. Single pixel wadi and rangetop one year time series (top) demonstrate similar annual cycles, but with varying rate of decay, timing, and amplitude. Spatial averages of the entire Jabal Qara (middle) contain features from both wadis and rangetops. Substantial interannual variability is clearly present in timing and amplitdue of phenology. Dashed line corresponds to 17-year mean of the spatial average. Average 17 year NDVI time series for the Qara (bottom) relative to its mean shows the aggregate phenology for the entire range across all years. Cumulative sum of this average shows a clear minimum at the end of the khareef as well as a gradual departure from subhorizontal at the beginning of the khareef, indicating the average beginning and ending dates of the khareef (red).**

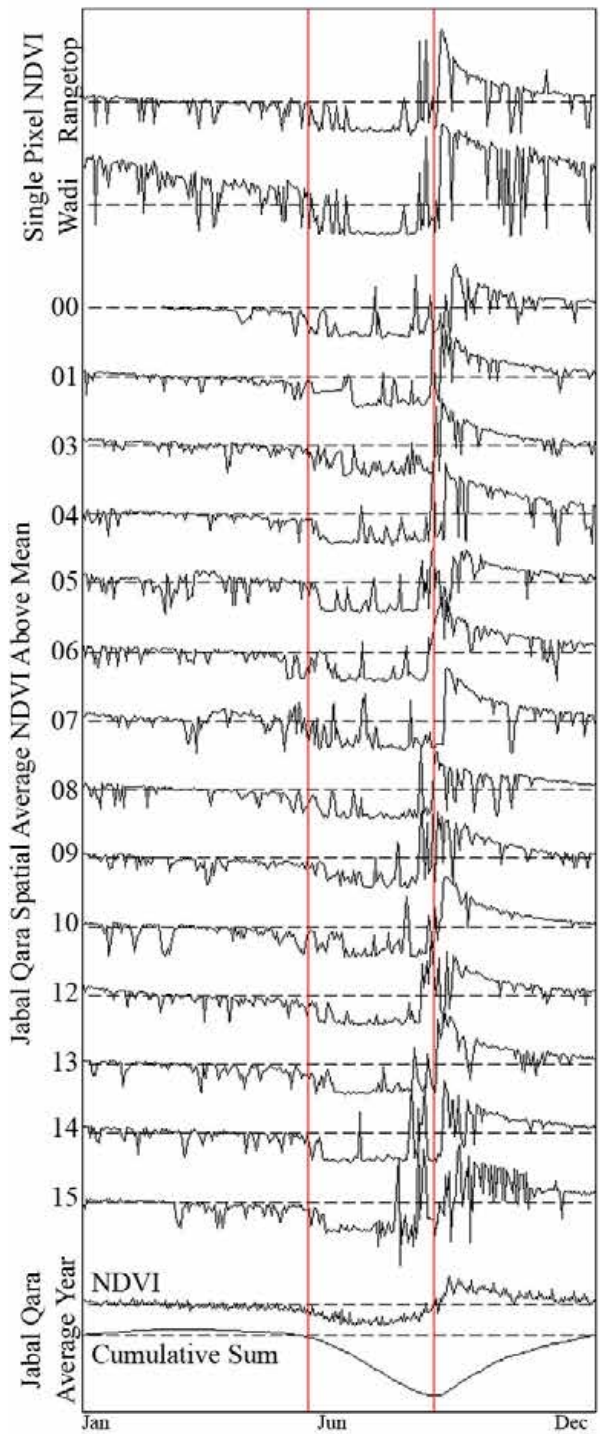

Figure 4. Interannual comparison of annual NDVI time series. Single pixel wadi and rangetop one year time series (top) demonstrate similar annual cycles, but with varying rate of decay, timing, and amplitude. Spatial averages of the entire Jabal Qara (middle) contain features from both wadis and rangetops. Substantial interannual variability is clearly present in timing and amplitdue of phenology. Dashed line corresponds to 17-year mean of the spatial average. Average 17 year NDVI time series for the Qara (bottom) relative to its mean shows the aggregate phenology for the entire range across all years. Cumulative sum of this average shows a clear minimum at the end of the khareef as well as a gradual departure from subhorizontal at the beginning of the khareef, indicating the average beginning and ending dates of the khareef (red).

The complement to the image of the temporal mean shown in Figure 3 is a time series of the spatial mean. The average NDVI across the whole of the Jabal Al Qara is shown for each year (except 2002, 2011 and 2016) in Figure 4. Examination of these curves shows the spatially coherent aggregate behavior of the Jabal Al Qara throughout the length of the dataset, revealing both year-to-year consistency and substantial interannual variability in the timing of the monsoon cloud cover and amplitude of the overall greening. The years 2002, 2011, and 2016 were excluded from this plot because of problems with data quality (discussed below).

The grand mean (mean NDVI of the entire dataset across both space and time) is shown as a dashed horizontal line on each plot in Figure 4. The very low NDVI values of the khareef generally plot below the grand mean, and the remainder of the year is generally above the mean. An abrupt crossing of the mean occurs at the end of the khareef, and a more gradual crossing of the grand mean occurs at the beginning of the khareef. The rate of change of the spatial mean indicates the spatial consistency of cloud cover over the Qara. More abrupt changes suggest abrupt transitions between widespread cloud and generally clear conditions.

The bottom of Figure 4 shows the mean-removed time series for the average year. After subtracting the grand mean, this time series is now partitioned into positive values (greater than grand mean) and negative values (lesser than grand mean). The cumulative sum of the mean-removed time series is also shown. The cumulative sum operator confers the advantage of effectively filtering out high frequency variability while faithfully preserving the long-timescale pattern of cloud cover. This can be used as a robust metric to quantify the ending of the khareef. The onset of the khareef is less abrupt, and thus more difficult to reliably determine – which is an important observation in and of itself. The use of the cumulative sum of the mean-subtracted data as a metric for monsoon parameters will be further explored later in this analysis.

*Spatiotemporal Patterns and Dimensionality*

**Figure 5. Spatial correlation, covariance, and partition of variance for daily MODIS NDVI image time series of the Jabal Qara and Jabal Qamar. Fused covariance (upper triangles) and correlation matrices (lower triangles) show low values during the khareef because of cloud cover. Covariance and correlation increase abruptly after the khareef because of extensive, near uniform, greening and low cloud cover. Substantial interannual variability is present in the beginning and ending of khareef, maximum greenness, and post-khareef senescence. Variance is partitioned comparably for both covariance (black) and correlation (red) based transforms. Spatial mean NDVI (dimension 1) variance is relatively higher for the Qamar, suggesting greater relative temporal variability for the Qara.**

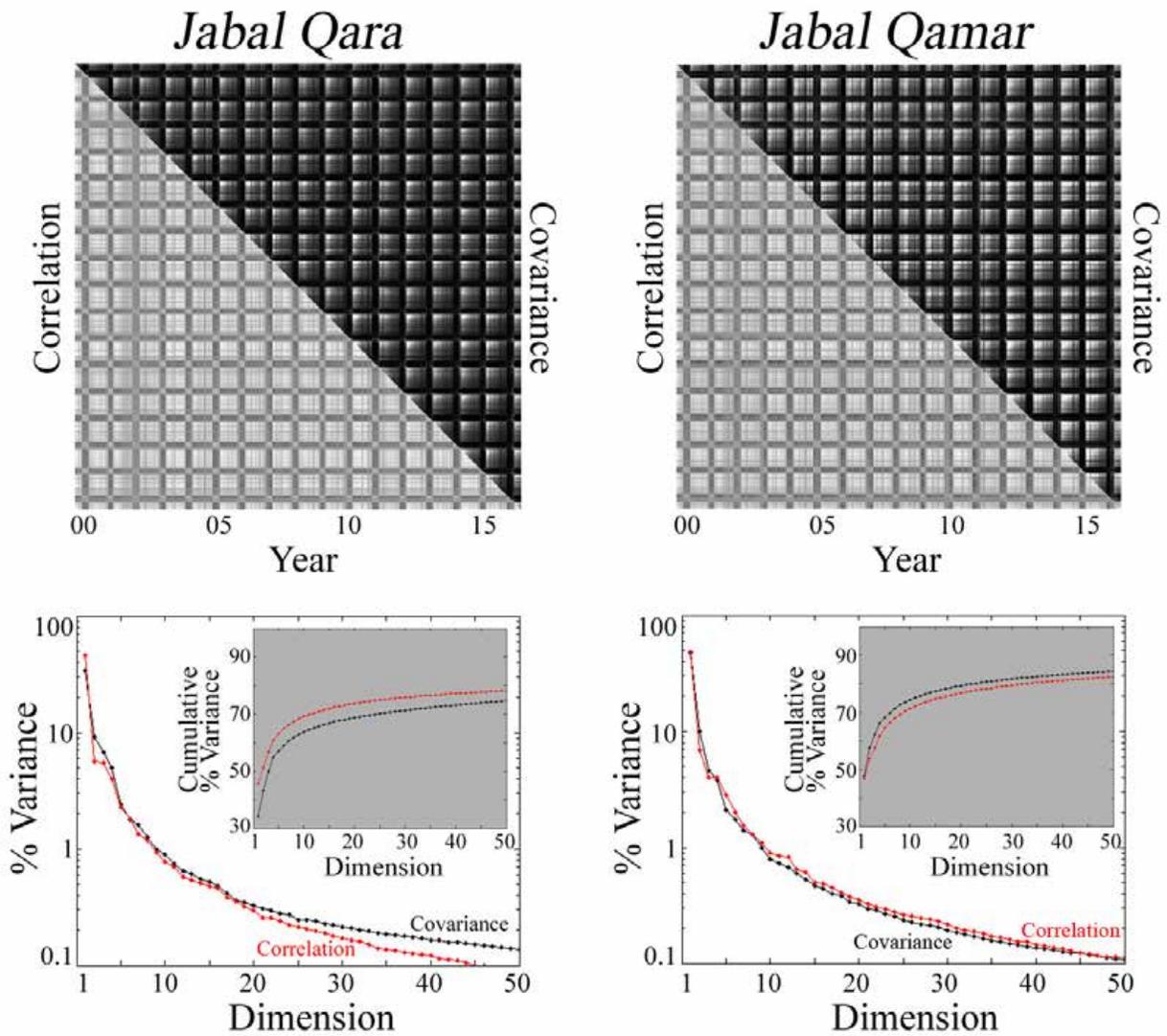

Figure 5. Spatial correlation, covariance, and partition of variance for daily MODIS NDVI image time series of the Jabal Qara and Jabal Qamar. Fused covariance (upper triangles) and correlation matrices (lower triangles) show low values during the khareef because of cloud cover. Covariance and correlation increase abruptly after the khareef because of extensive, near uniform, greening and low cloud cover. Substantial interannual variability is present in the beginning and ending of khareef, maximum greenness, and post-khareef senescence. Variance is partitioned comparably for both covariance (black) and correlation (red) based transforms. Spatial mean NDVI (dimension 1) variance is relatively higher for the Qamar, suggesting greater relative temporal variability for the Qara.

The spatial covariance and correlation matrices are another representation of aggregate spatiotemporal structure of the dataset. They also form the basis for the Empirical Orthogonal Function (EOF) transform used in the next portion of the analysis.

Each cell of these matrices shows the correlation or covariance between the spatial patterns of NDVI at the time of the row number and the time of the column number. Bright cells represent pairs of dates with high similarity in the spatial structure and/or amplitude of their NDVI patterns. Dark cells represent essentially uncorrelated spatial structures and/or low amplitude overall NDVI signals. Although it is possible for spatial patterns to be negatively correlated, this is not observed in this dataset.

The correlation and covariance matrices show both annual periodicity and interannual variability in both the Jabal Al Qara and Jabal Al Qamar. In both ranges, the khareef is clearly present as an annual dark band of decorrelation in both the covariance and correlation matrices, as clouds are characterized by generally near-zero NDVI values and rapidly variable spatial patterns. The post-khareef vegetation phenology is characterized as an abrupt brightening in the covariance matrix as the clouds clear to reveal a spatially stable landscape of dense, photosynthetic vegetation with high NDVI values. This peak in covariance is followed by rapid decay due to simultaneous grazing and senescence. This decay is more prominent in the Qara than the Qamar due to the larger relative area of rapidly senescing rangetop grasslands. This decay is not prominent in the correlation matrix because the spatial pattern of post-khareef vegetation remains essentially constant, despite the fluctuations in overall amplitude of greenness.

Both ranges show similar annual cycles, as expected from regions dominated by similar large-scale climatic and ecological processes. However, differences between the two ranges are also visible. The amplitude and timing of the post-khareef senescence is clearly different between the two ranges and variable from year to year.

We use EOF analysis [*Lorenz*, 1959] to find orthogonal basis functions for the > 4000 dimensional image cubes. This technique leverages the band-to-band covariance (or correlation) to decompose the data matrix into uncorrelated modes of variability (*dimensions*) using a Principal Component (PC) transformation. Each dimension contains an associated spatial and temporal pattern, and each dimension is uncorrelated with every other dimension. Every dimension also represents some fraction of the total variance of the dataset. After EOF analysis, the transformed dataset has the same number of dimensions as the original dataset, and the total variance remains the same, but the variance has been partitioned in such a way to compress the maximum fraction of the total variance into the minimum number of uncorrelated dimensions. EOF analysis can be performed with total variance of each image normalized (correlation) or unnormalized (covariance), with slightly different results. For an excellent introduction to EOF analysis, see [*von Storch and Zweirs*, 2002].

The bottom portion of Figure 5 shows the partition of variance from the EOF analysis for the Jabal Al Qara and Jabal Al Qamar data matrices. Most prominently, the Jabal Al Qamar shows substantially more variance in the low order dimensions than the Jabal Al Qara. In addition, subtle differences in the partition of variance between the correlation and covariance based transforms are present in both ranges. While detailed comparison of these differences is an interesting subject, it is beyond the scope of this analysis.

We choose to work with the covariance-based transform for the remainder of this analysis. As seen in Figure 5, the spatial covariance matrix contains more of the amplitude structure of the vegetation phenology than the correlation matrix, central to the objectives of this study. Furthermore, all images contain measurements of the same normalized quantity (NDVI), so further band-to-band normalization is not strictly necessary.

*Characterization of Spatiotemporal Patterns with the Temporal Feature Space*

Characterization is crucial for spatiotemporal analysis. The complexity of many spatiotemporal processes suggests that detailed characterization can provide insight for understanding the system and informing subsequent modeling. In some cases, characterization of a complex spatiotemporal dataset can yield simple, informative structure which may not be otherwise apparent. This structure can give novel insight into the processes under study. Empirical Orthogonal Function (EOF) analysis [*Lorenz*, 1959] is a common, well-understood method for high dimensional spatiotemporal data analysis [*Eshel*, 2012]. When combined with temporal mixture modeling and Fourier analysis, EOF analysis can provide a robust, straightforward spatiotemporal approach to characterizing seasonal and interannual processes in time series of synoptic imagery [*Small*, 2012]. In short, this approach allows us to simultaneously characterize the seasonal and interannual timing, duration and periodicity of both cloud cover and vegetation phenology from a single time series of daily synoptic satellite observations without the need for *a priori* assumptions about the functional form of the cycles we seek to quantify.

The low-dimensional representation of a high-dimensional remote sensing data set is generally referred to as a *feature space* [*Landgrebe*, 1973]. A spectral feature space represents the diversity of pixel reflectance spectra in an image as a scatterplot in which pixels with similar spectra plot closer together and pixels with more distinct spectra plot further apart. The temporal analog, referred to as a temporal feature space (TFS), represents the diversity of temporal patterns in an analogous way, in which the partition of variance (given by the EOF analysis) gives an indication of the dimensionality of the TFS [*Small*, 2012]. Following the approach of [*Small*, 2012], we use the spatial PCs of the first 3 dimensions of the PC transform to represent the low order structure of the temporal feature space for the 17 year MODIS data cube (Figure 6). In these scatterplots, every point represents the time series of an image pixel (or many pixels). The position of the pixel in this temporal feature space is determined by the relative contribution of the first three uncorrelated modes of variability.

**Figure 6a. Jabal al Qara temporal feature space. Only the first 3 dimensions are shown. The PC 3/2 projection of the TFS is clearly partitioned between wadis (bottom) and grasslands (top). In this projection, longitudinal gradations are also present. Eastern Rangetops (ER) grade through Central Rangetops (CR) to the Western Rangetops (WR) from left to right. Wadis also grade from Eastern Wadis (EW) to Western Wadis (WW) from left to right. Western wadis are more clearly distinct on the PC 1/2 projection since they overprint the Nejd in the PC 3/2 projection. The Nejd EM is analogous to the origin of the feature space because it has almost no variance. Endmember mean time series for 1001 pixels chosen at corresponding corners of the TFS shown in the lower right corner.**

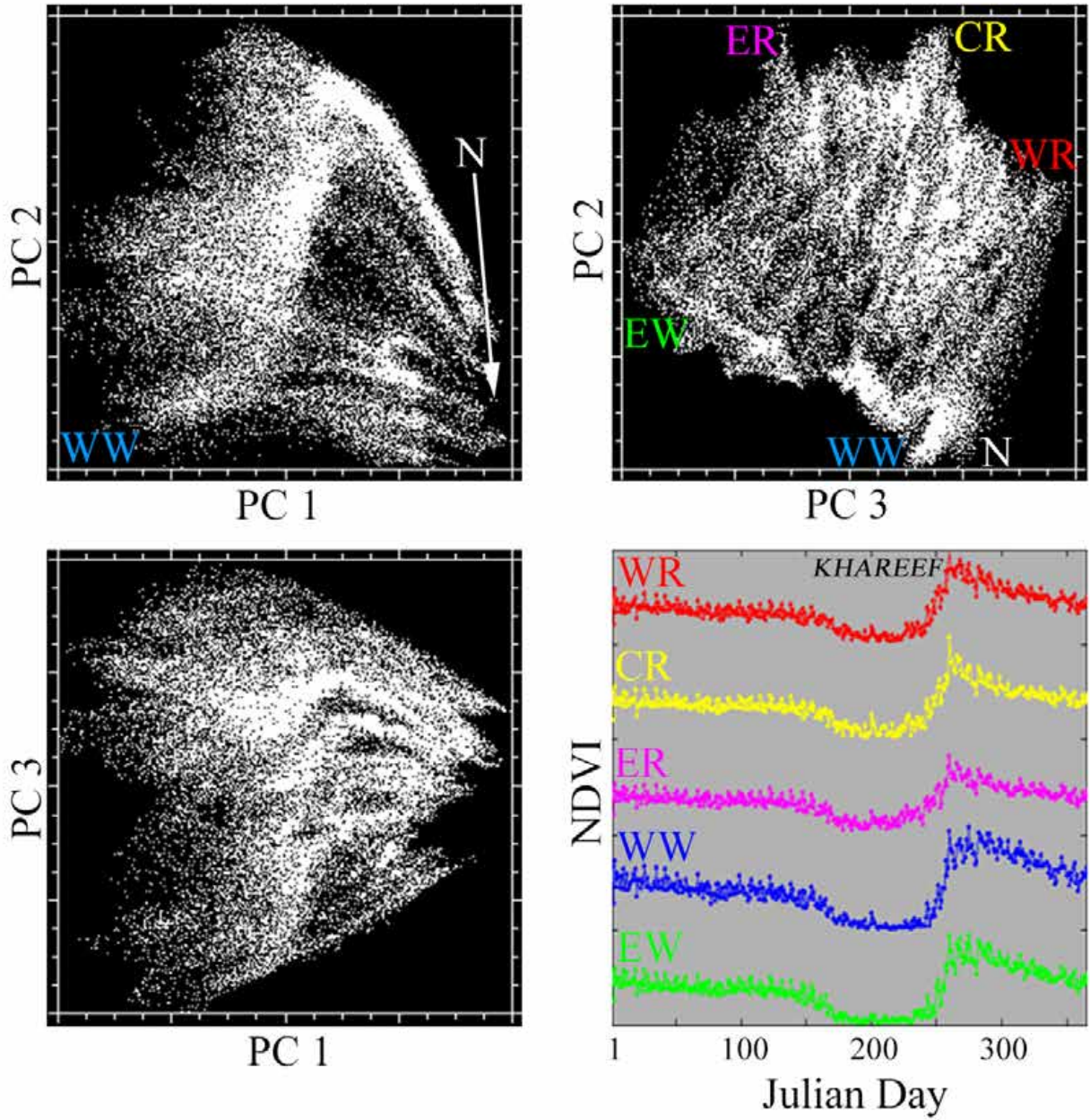

Figure 6a. Jabal al Qara temporal feature space. Only the first 3 dimensions are shown. The PC 3/2 projection of the TFS is clearly partitioned between wadis (bottom) and grasslands (top). In this projection, longitudinal gradations are also present. Eastern Rangetops (ER) grade through Central Rangetops (CR) to the Western Rangetops (WR) from left to right. Wadis also grade from Eastern Wadis (EW) to Western Wadis (WW) from left to right. Western wadis are more clearly distinct on the PC 1/2 projection since they overprint the Nejd in the PC 3/2 projection. The Nejd EM is analogous to the origin of the feature space because it has almost no variance. Endmember time series for 1001 pixels chosen at corresponding corners of the TFS shown in the lower right corner.

**Figure 6b. Jabal al Qamar temporal feature space. Only the first 3 dimensions of the covariance based transform are shown. The TFS is again clearly partitioned, but this time most usefully in the PC 1/2 projection. Wadis (right) and Nejd (left) are clearly distinct. Longitudinal gradations are present again, but this time in dimension 2. Western Wadis (WW) grade to Eastern Wadis (EW). Central wadis are not distinct enough to form an additional EM. Eastern and Western Nejd pixels clearly grade on a continuum toward a single null EM. However, the spatial domain of the analysis does not extend deeply into the Nejd so east and west are treated separately. The PC 3/2 projection forms a phase plane with Nejd pixels at the center and wadi pixels on the circumference. EM time series for 101 pixels chosen at corresponding corners of the TFS shown in the lower right.**

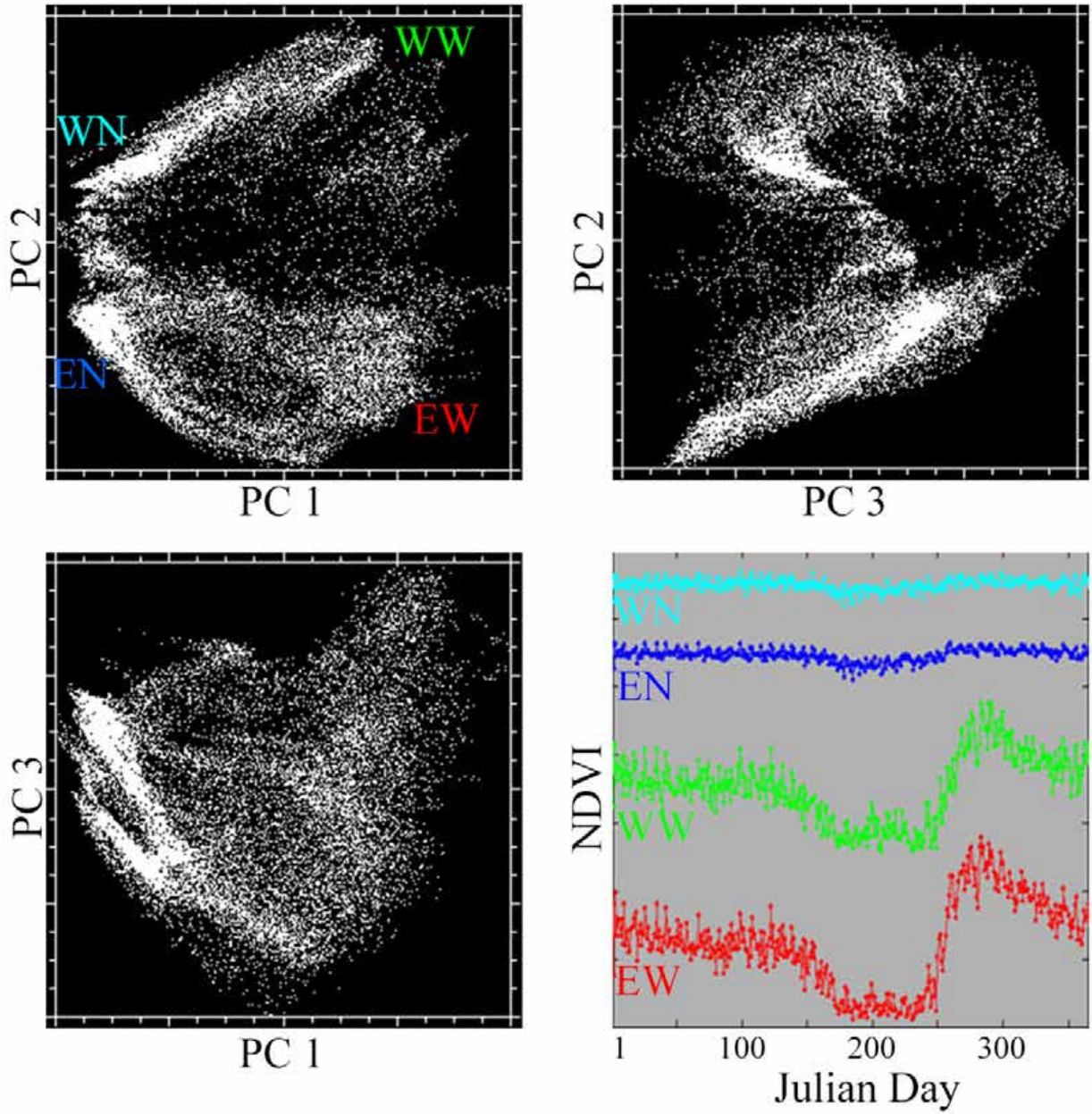

Figure 6b. Jabal al Qamar temporal feature space. Only the first 3 dimensions of the covariance based transform are shown. The TFS is again clearly partitioned, but this time most usefully in the PC 1/2 projection. Wadis (right) and Nejd (left) are clearly distinct. Longitudinal gradations are present again, but this time in dimension 2. Western Wadis (WW) grade to Eastern Wadis (EW). Central wadis are not distinct enough to form an additional EM. Eastern and Western Nejd pixels clearly grade on a continuum toward a single null EM. However, the spatial domain of the analysis does not extend deeply into the Nejd so east and west are treated separately. The PC 3/2 projection forms a phase plane with Nejd pixels at the center and wadi pixels on the circumference. EM time series for 101 pixels chosen at corresponding corners of the TFS shown in the lower right.

The geometrical structure of the Jabal Al Qara temporal feature space (Figure 6a) is characterized by corners and sharp edges. The pixel time series at the corners are the endmembers (EMs) of the feature space, representing the statistically most distinct temporal patterns. The continuum bounded by the EMs is indicative of gradations among the extreme temporal patterns associated with pixels at the corners and edges. The sharp edges suggest that most pixels in the image can be well represented by combinations of a relatively small number of EM temporal patterns.

Six distinct EMs are identifiable from the Jabal Al Qara feature space. The EMs are shown in the lower right while their locations are shown on the PC scatterplots. Most of these corners are most clearly visible from the projection showing dimensions 2 vs 3 (upper right). Some corners are obscured from ready identification in the other projections since they plot above or below the plane of the page and are therefore projected onto one another in a 2D scatterplot.

The two EMs which are most distinct from the dimension 1 vs 2 scatterplot are the Null and Western Wadi EMs. The Null EM corresponds to pixels with essentially no vegetation and is analogous to the origin of the feature space. The western wadis, on the other hand, show the highest amplitude phenology of all the temporal patterns in the Qara. The ready identification of these two patterns using dimensions 1 and 2 is intuitive since the first dimension of an EOF transform corresponds to the temporal mean NDVI. These EMs plot on top of each other in the bottom of the dimension 3 vs 2 scatterplot, from which all the other EMs are identified.

The other EMs form clear corners on the scatterplot of dimensions 3 vs 2. This combination of dimensions is fortuitous as its geometry corresponds nearly directly to the geography of the Qara. The wadis and rangetops are clearly distinct from each other along an axis oblique to either of the individual dimensions. Furthermore, the eastern and western portions of the ranges are distinct from each other along an axis nearly perpendicular to the wadi-rangetop axis. This results in a boxy character to the dimension 3 vs 2 scatterplot.

Individual wadi systems are identifiable from the temporal feature space. Each curvilinear collection of pixels grading from the lower edge of the dimension 3 vs 2 projection to the upper edge corresponds to a spatially coherent set of pixels from the base of an individual wadi up through higher elevations to the rangetops immediately surrounding it. The full Jabal Al Qara region is thus composed of distinct phenologic systems which grade as a continuum from east to west. These individual wadis are also present radiating outward from the null EM in the dimension 1 vs 2 and 1 vs 3 projections.

Samples of 1001 pixel time series were chosen from each of the EM regions in the temporal feature space. In every case, these pixels cluster in geographic space to form EM regions which are nearly spatially contiguous. The pixel time series for each region were then averaged together to form one mean EM time series representative of that spatiotemporally distinct region. The 5 averages of these 5005 pixel time series concisely represent the spatiotemporal variability of all 33,729 pixel time series in the Jabal al Qara subset of the NDVI image stack.

The time series for each of the EMs is shown in the lower right of Figure 6. For display purposes, the 17 years for each EM time series were again averaged together to form a single representative year. The full 17 time series for each EM region is discussed in detail below. While each statistically distinct region has a similar average annual cycle, variations in amplitude of post-khareef greenness and rate of senescence are clearly visible. These variations will be discussed below.

The temporal feature space for the Jabal Al Qamar (Figure 6b) exhibits substantially less complexity than the Jabal Al Qara. While the wadis and rangetops are again distinct, the minimal spatial extent of the rangetops results in the feature space grading directly from forested wadis into the sparsely vegetated Nejd. While two distinct Nejd corners are evident in the dimension 1 vs 2 space, these are clearly grading toward a single Null EM. Had the spatial extent of the analysis been expanded further inland, these continua would merge together.

Examination of the dimension 1 vs 2 projection also demonstrates that, while a continuum does exist between the eastern and western wadis of the Qamar, the temporal patterns are much more clustered in space than they are in the Qara, with a narrower, more abrupt transition zone.

*Aside: Comments on Spatiotemporal Analysis using the Temporal Feature Space*

Using the temporal feature space comprised of low order EOFs to identify the dominant spatiotemporal patterns in the dataset prevents the common mistake of overinterpreting the physical significance of individual EOFs (Small, 2012). Because each pixel time series is represented as a linear combination of all the individual EOFs, which interfere both constructively and destructively, a single EOF rarely represents the majority of the structure of a single time series. The temporal feature space shows graphically how the low order EOFs combine to represent the dominant features of each pixel time series. While over interpreting a single EOF can be misleading, the relative contributions of the low order EOFs shown in the feature space can show how the patterns of these individual EOFs combine to represent the important temporal features present in the dataset. With this approach we avoid over interpretation of individual EOFs and merely use them to a) quantify the dimensional complexity of the dataset and b) geometrically arrange the data along the dimensions of maximum variance. In other words, individual EOFs rarely act alone. The temporal feature space shows how different EOFs combine to represent the most statistically distinct temporal patterns in the dataset.

While the first three dimensions alone are inadequate to describe all of the complexity of the dataset, they clearly yield sufficient information about the spatiotemporal patterns of this particular dataset to provide insight into the nature of the system. We suggest that limiting the current investigation to the low-order dimensions of the space is justified because of the substantial new information it presents and the relative geometrical simplicity of the interpretation. Furthermore, the effectively random variance present in the data results in substantial variance being partitioned to high dimensions (and thus removed from a low-order analysis). However, a more complete characterization involving higher dimensions of the feature space could yield further insight into the system.

*Maps of Distinct Spatiotemporal Patterns*

**Figure 7. Temporal affinity maps and endmember regions for Jabal al Qara (top) and Jabal al Qamar (bottom). A. Wadi vegetation phenology in the Qara is largely divided between distinct east and west regions with some mixing in the central portion of the range. B. Rangetop vegetation phenology in the Qara exists on a continuum with higher maximum greenness and slower senescence in the west than the east. C and D. Wadi phenology in the Qamar is also divided by longitude into two distinct clusters, with limited mixing in the central portion of the range. While both Nejd Qara EMs are essentially null, enough statistical difference, predominantly in patterns of cloud cover, exists to separate them into**

*two distinct groups. All colors are stretched linearly from 0% to 100% affinity. EM locations are outlined in gray.*

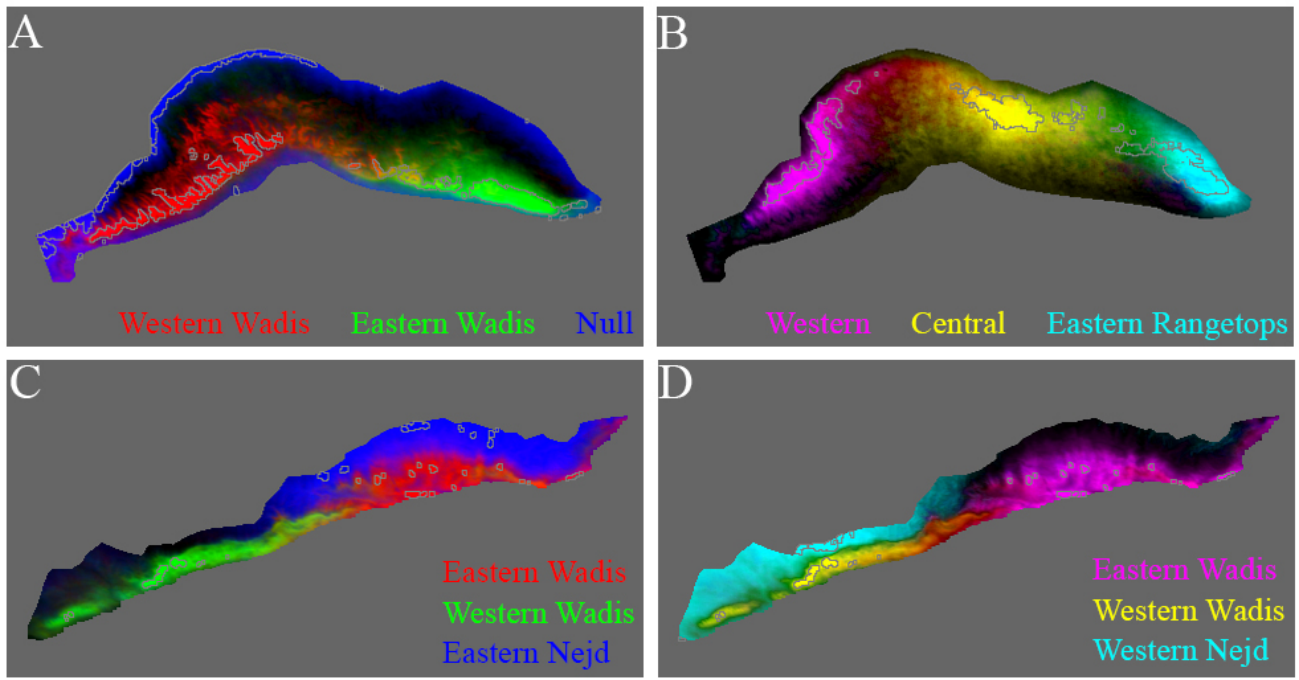

Figure 7. Temporal affinity maps and endmember regions for Jabal al Qara (top) and Jabal al Qamar (bottom). A. Wadi vegetation phenology in the Qara is largely divided between distinct east and west regions with some mixing in the central portion of the range. B. Rangetop vegetation phenology in the Qara exists on a continuum with higher maximum greenness and slower senescence in the west than the east. C and D. Wadi phenology in the Qamar is also divided by longitude into two distinct clusters, with limited mixing in the central portion of the range. While both Nejd Qara EMs are essentially null, enough statistical difference, predominantly in patterns of cloud cover, exists to separate them into two distinct groups. All colors are stretched linearly from 0% to 100% affinity. EM locations are outlined in gray.

The topology of the temporal feature space provides a concise representation of the temporal EMs bounding the space and the distribution of temporal patterns within the space. This continuum of temporal patterns can be represented geographically with a temporal mixture model. Temporal mixture models represent each pixel time series as a linear combination of the temporal EMs identified from the temporal feature space [*Piwowar et al.*, 2006]. The correspondence between the statistically derived basis functions (the EOFs) and the physical basis functions (the temporal EMs) is analogous to a transformation that allows us to interpret the temporal patterns in each pixel time series in terms of their temporal EMs rather than their EOFs alone (Small, 2012). We use a temporal mixture model to represent all pixels in the data set as linear combinations of the 1001 (or 101) pixel EM averages identified from the Jabal Al Qara (or Qamar) feature space. Each pixel time series is thus characterized by a set of 6 (or 4) weights, plus an error term quantifying model misfit. For each of these models, 95% of all pixel misfits were below 5%.

Maps of these temporal mixture models are shown in Figure 7. In each of these tricolor composite maps, three primary colors (RGB or CMY) represent the weights of three of the statistically distinct temporal EMs identified from the feature space of Figure 6. Because there are more than 3 EM regions in each mountain range, a single image is insufficient to display all 4 (or 6) of the temporal EM fractions. Fractions are displayed linearly in the range of 0 to 1 for each EM. For instance, in Panel A, a pixel in the western wadis of the Qara would appear as saturated red (100% red, 0% green, 0% blue), while a more centrally located wadi pixel could appear as yellow (50% red, 50% green, 0% blue). In contrast, a Qara grassland pixel would appear as black (0% red, 0% green, 0% blue) in Panel A since no grassland EMs are represented in that map, but bright in Panel B. In each panel, the locations of the EM pixels selected from the temporal feature space are outlined in gray.

A principal finding of this work is that the temporal patterns cluster strongly in space. This is evident in the locations of EM pixels both ranges (particularly the Qara) as well as the spatial distribution of these temporal patterns in non-EM locations. This spatial coherence emerges directly from the structure of the dataset - without any prescribed assumptions about the geography or timing of the monsoon or phenology. The geographic consistency of the EM pixels selected from the apexes of the temporal feature space is an important result because it reinforces the notion that structure of the feature space is physically meaningful and that the temporal patterns in the feature space are related to geographic factors like elevation, slope and coastal proximity. The spatial coherence of the EM pixels was not imposed *a priori*, and is not required by the analysis methods used.

The degree of clustering of the EM locations is particularly prominent in the western wadi cloud forests of the Qara (Panel A, red). The most distinct temporal patterns clearly track individual wadis and do not include the rangetop grasslands separating them. The eastern wadis (green) also form a spatially coherent cluster distinct from the western wadis, with little mixing between the two. The spatial extent of regions with minimal vegetation (blue) is most prominent in the Nejd but also includes barren regions of the coastal plain.

In contrast, the vegetation phenology of the rangetop grasslands of the Jabal Al Qara (Panel B) shows a remarkably smooth gradation between distinct western (magenta), central (yellow) and eastern (cyan) patterns. Notably, the phenology of the grasslands in the central Qara emerges as a distinct pattern which is not a simple binary mixture of the eastern and western grassland EMs.

The vegetation phenology in the Jabal Al Qamar (Panels C and D) also shows clear east-west differentiation. The phenology of the wadis is clearly broken into an eastern and a western zone, with a rather sharp gradient. Spatial mixing in the Qamar is likely minimized by the rugged topography which physically divides the range. While the Nejd also shows strong east-west variation in space, the temporal patterns of each EM region (shown below, Figure 8b) are so similar to each other, and contain so little signal, that conservatively we do not conclude any significant phenological distinction between the two regions. As mentioned above, both Nejd EMs clearly grade into a single null EM that would encompass both regions if the study area were expanded further inland.

*Time Series of Distinct Spatiotemporal Patterns*

Time series of the distinct spatiotemporal patterns provide information complementary to the spatial patterns shown above. We present these time series as spatial averages of the EM regions identified from the temporal feature space in Figure 6 and outlined in gray in Figure 7.

In order to demonstrate both the overall phenological signal and the interannual variability in each distinct EM region, we display the annual mean +/- 1 standard deviation for each EM region (Figure 8). To examine individual years, we then display the full 17-year time series for each EM region (Figure 9).

**\*\*Figure 8a. Annual cycle of Jabal Qara EM time series. Mean (black) and +/- 1 standard deviation (red) shown for each Julian Day for daily NDVI observations from 2000-2017. Cloud forests generally reach higher maximum NDVI values and senesce slower than rangelands. Western regions of both rangetops and cloud forests senesce slower and show more variability than eastern regions. Clouds clear earlier over rangetops than wadis and are more consistent in west than east. Irregular breaks in clouds during the khareef are more prevalent in rangetops than wadis. On average, onset of the khareef is more gradual than termination, especially in the cloud forests. Nejd values are low year-round, with a low amplitude but consistent annual cycle due to regular variations in solar geometry. Annual means for each EM region are shown as thin horizontal lines.**

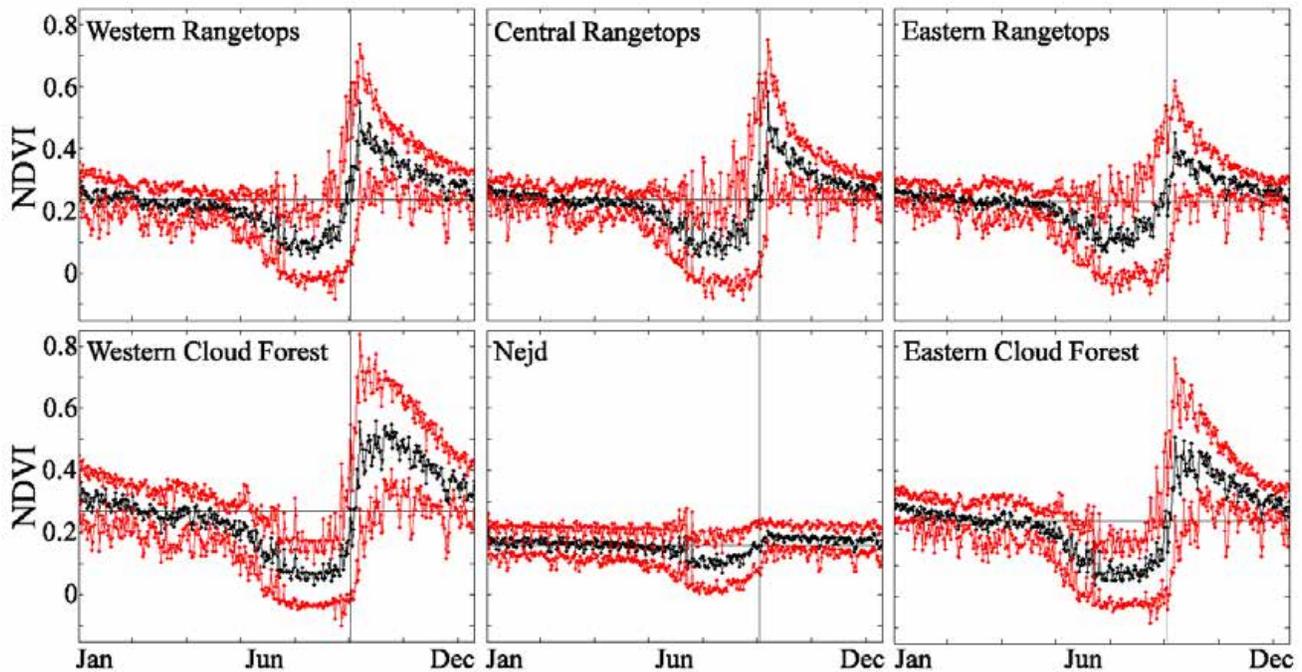

Figure 8a. Annual cycle of Jabal Qara EM time series. Mean (black) and +/- 1 standard deviation (red) shown for each Julian Day for daily NDVI observations from 2000-2017. Cloud forests generally reach higher maximum NDVI values and senesce slower than rangelands. Western regions of both rangetops and cloud forests senesce slower and show more variability than eastern regions. Clouds clear earlier over rangetops than wadis and are more consistent in west than east. Irregular breaks in clouds during the khareef are more prevalent in rangetops than wadis. On average, onset of the khareef is more gradual than termination, especially in the cloud forests. Nejd values are low year-round, with a low amplitude but consistent annual cycle due to regular variations in solar geometry. Annual means for each EM region are shown as thin horizontal lines. Time of minimum cumulative NDVI for the entire region is shown as a thin vertical bar for reference.

**Figure 8b. Annual cycle of Jabal Qamar EM time series. Mean (black) and +/- 1 standard deviation (red) shown for each Julian Day for daily NDVI observations from 2000-2017. Opposite from the Qara, cloud forests in the Qamar generally have higher maximum values and slower senescence in the east than the west. In some wet years, regreening post-kharif is present in both the Qara and Qamar. This is prominent in both western and eastern Qamar EM time series. Similar to the Qamar, onset of the khareef is more gradual than termination. Nejd values are low year-round. East-west variations in Nejd EMs are barely visible, predominantly due to variations in cloud cover during the Khareef. The low amplitude cycle due to solar illumination is also faintly present. Annual means for each EM region are shown as thin horizontal lines.*

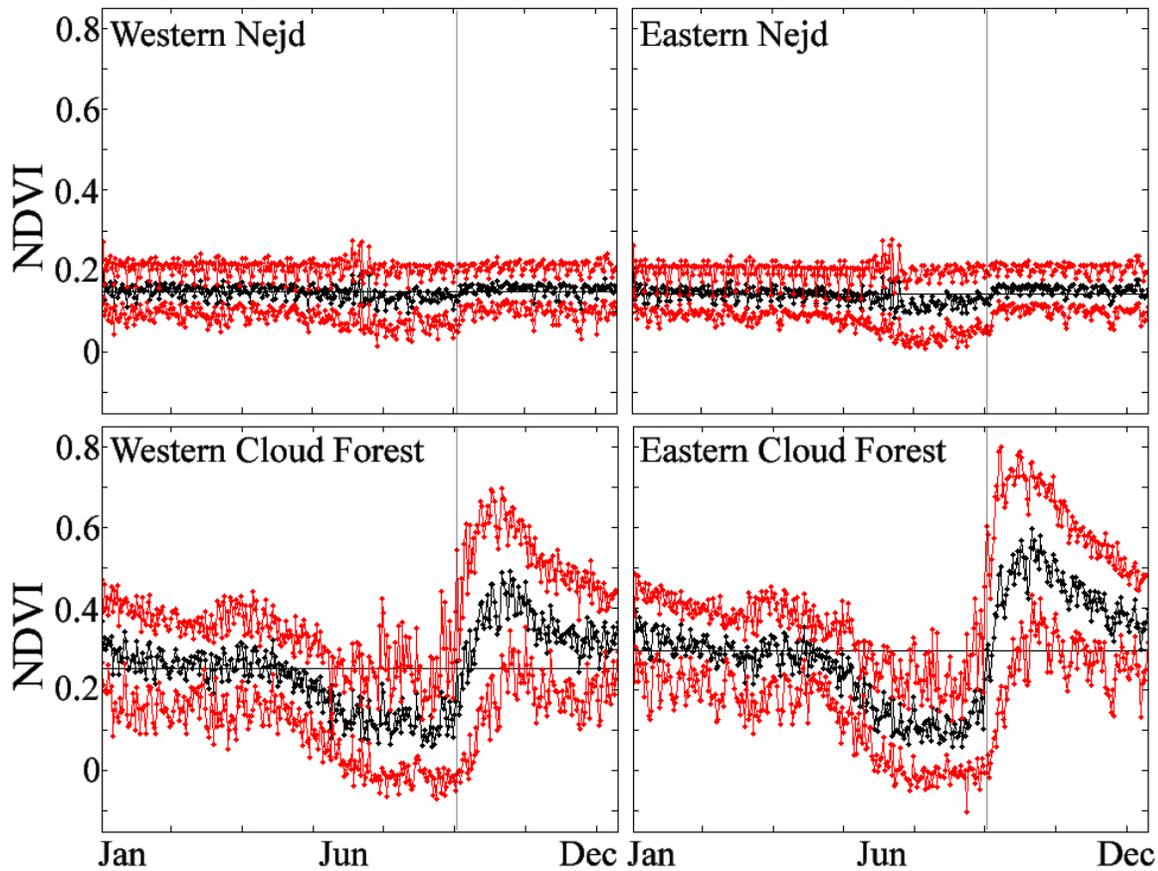

Figure 8b. Annual cycle of Jabal Qamar EM time series. Mean (black) and +/- 1 standard deviation (red) shown for each Julian Day for daily NDVI observations from 2000-2017. Opposite from the Qara, cloud forests in the Qamar generally have higher maximum values and slower senescence in the east than the west. In some wet years, regreening post-kharif is present in both the Qara and Qamar. This is prominent in both western and eastern Qamar EM time series. Similar to the Qamar, onset of the khareef is more gradual than termination. Nejd values are low year-round. East-west variations in Nejd EMs are barely visible, predominantly due to variations in cloud cover during the Khareef. The low amplitude cycle due to solar illumination is also faintly present. Annual means for each EM region are shown as thin horizontal lines. Time of minimum cumulative NDVI for the entire region is shown as a thin vertical bar for reference.

Examination of time series of the phenology of the Jabal Al Qara EMs (Figure 8a) reveals both common behavior and substantial differences across geographical regions. The annual cycle in all cases is clearly dominated by the khareef, with consistently below average NDVI from mid-June through mid-September in all EMs, including the Nejd. As explained above, this negative NDVI anomaly results from the near-zero spectral slope of clouds compared to the small but positive spectral slope of bare soil, regolith, and NPV.

Rangetop grasslands in the Jabal Al Qara are clearly differentiated from wadi cloud forests based on the amplitude and rate of senescence of the post-khareef greenness. Regardless of longitude, wadis generally reach higher peak NDVI values and take longer to senesce than grasslands. Close inspection of the curves during the khareef also shows the wadis to have longer, less interrupted cloud cover than the rangetops. This is further examined with cumulative sums in Figure 11.

Additionally, a clear longitudinal signal is present in both the wadi cloud forests and rangetop grasslands in the Qara. The greening signal is observed to be stronger in the western portion of the range and weaker in the eastern portion of the range. It is the combination of this longitudinal gradient and the wadi-rangetop differentiation that gives the temporal feature space of Figure 6 its structure.

The cloud forests of the Jabal Al Qamar (Figure 8b) demonstrate a similar east-west gradient in phenological amplitude – but in the opposite direction. The observation that these gradients in phenological amplitude seem to converge in the unvegetated, high elevation region between the two ranges is interesting and could be the basis for further work.

In addition to the longitudinal gradient, both the eastern and western cloud forests of the Qamar show a prominent but variable early greening (or regreening) signal in the months immediately prior to the onset of the khareef. This is only weakly present in the annual averages of the cloud forests of the Qara, and will be shown in more detail in the full EM time series in Figure 9.

Substantially more variability is observed in the Qamar than the Qara throughout year. This is likely due to both more frequent non-khareef cloud cover and the fewer number of points included in the spatial average.

**Figure 9a. Full time series for spatial averages of 1001 pixel spectra from the Jabal Qara. Substantial variability is observed both between EMs and within each EM across years. Rangetops vary substantially from year to year in terms of maximum greenness, rate of senescence, and relative similarity of western, central, and eastern regions. Wadis also vary considerably from year to year. Some years show post-khareef regreening in one or both of the EM regions. Years with both strong and weak greening in both wadis and rangetops are present at the beginning and end of the time series. No clear trend in EM time series observed in terms of duration or timing of khareef, maximum greenness, or rate of senescence of any EM phenology.**

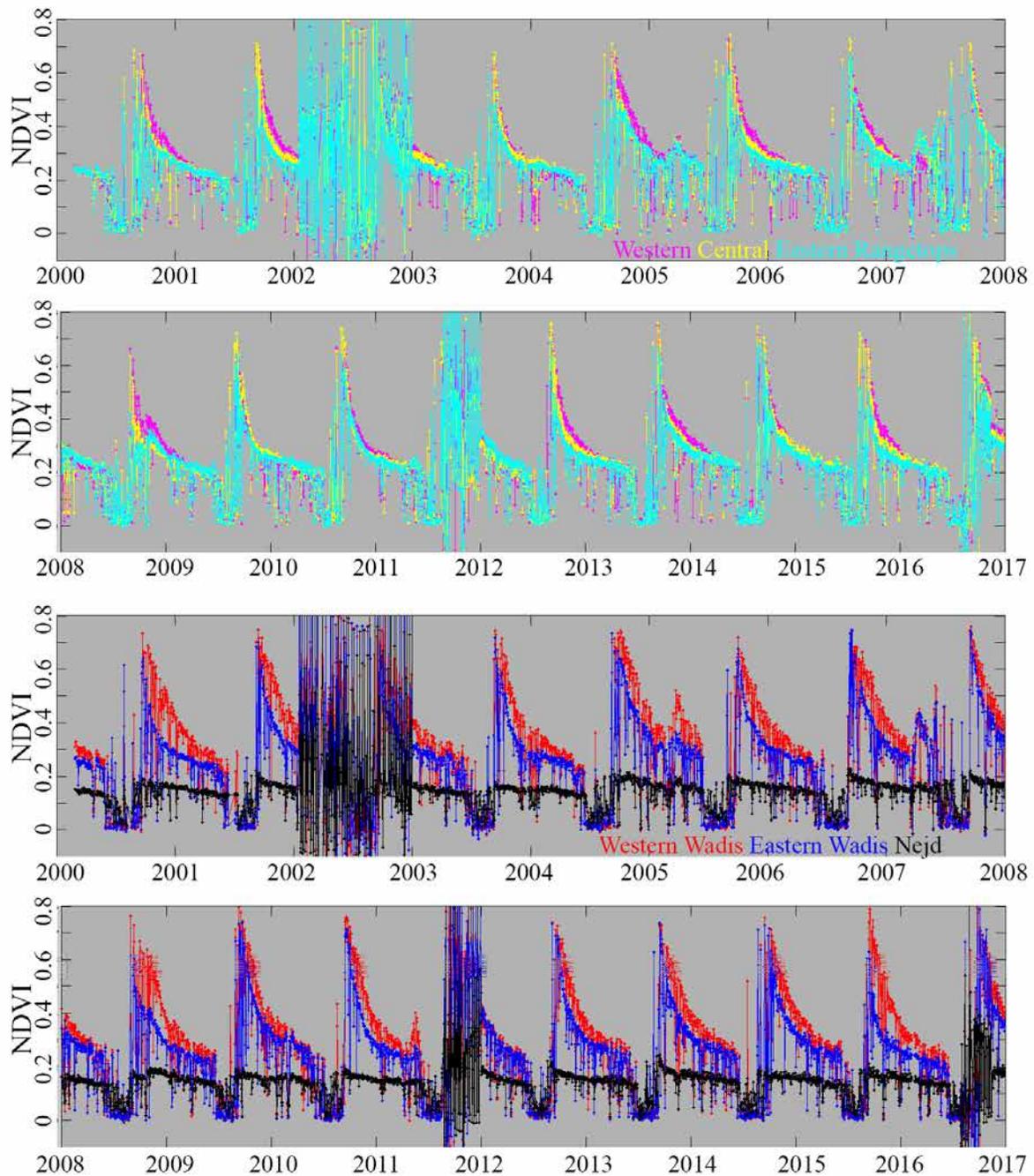

Figure 9a. Full time series for spatial averages of 1001 pixel spectra from the Jabal Qara. Substantial variability is observed both between EMs and within each EM across years. Rangetops vary substantially from year to year in terms of maximum greenness, rate of senescence, and relative similarity of western, central, and eastern regions. Wadis also vary considerably from year to year. Some years show post-khareef regreening in one or both of the EM regions. Years with both strong and weak greening in both wadis and rangetops are present at the beginning and end of the time series. No clear trend in EM time series observed in terms of duration or timing of khareef, maximum greenness, or rate of senescence of any EM phenology.

**Figure 9b. Full time series for spatial averages of 101 pixel spectra from the Jabal Qamar. As in the Qara, substantial variability is observed both between EMs and within each EM across years. Eastern wadis are generally greener than western wadis. Again, wadis vary considerably from year to year. Post-khareef regreening appears more frequent and more intense in the Qamar than the Qara. As in the Qara, years with both strong and weak greening of the Qamar are present at the beginning and end of the time series. Again, no clear trend in EM time series observed in terms of duration or timing of khareef, maximum greenness, or rate of senescence of either EM phenology. Minor but consistent differences are also present in the eastern and western Nejd. The values of NDVI involved are so low that we do not attempt to ascribe physical meaning to these differences.*

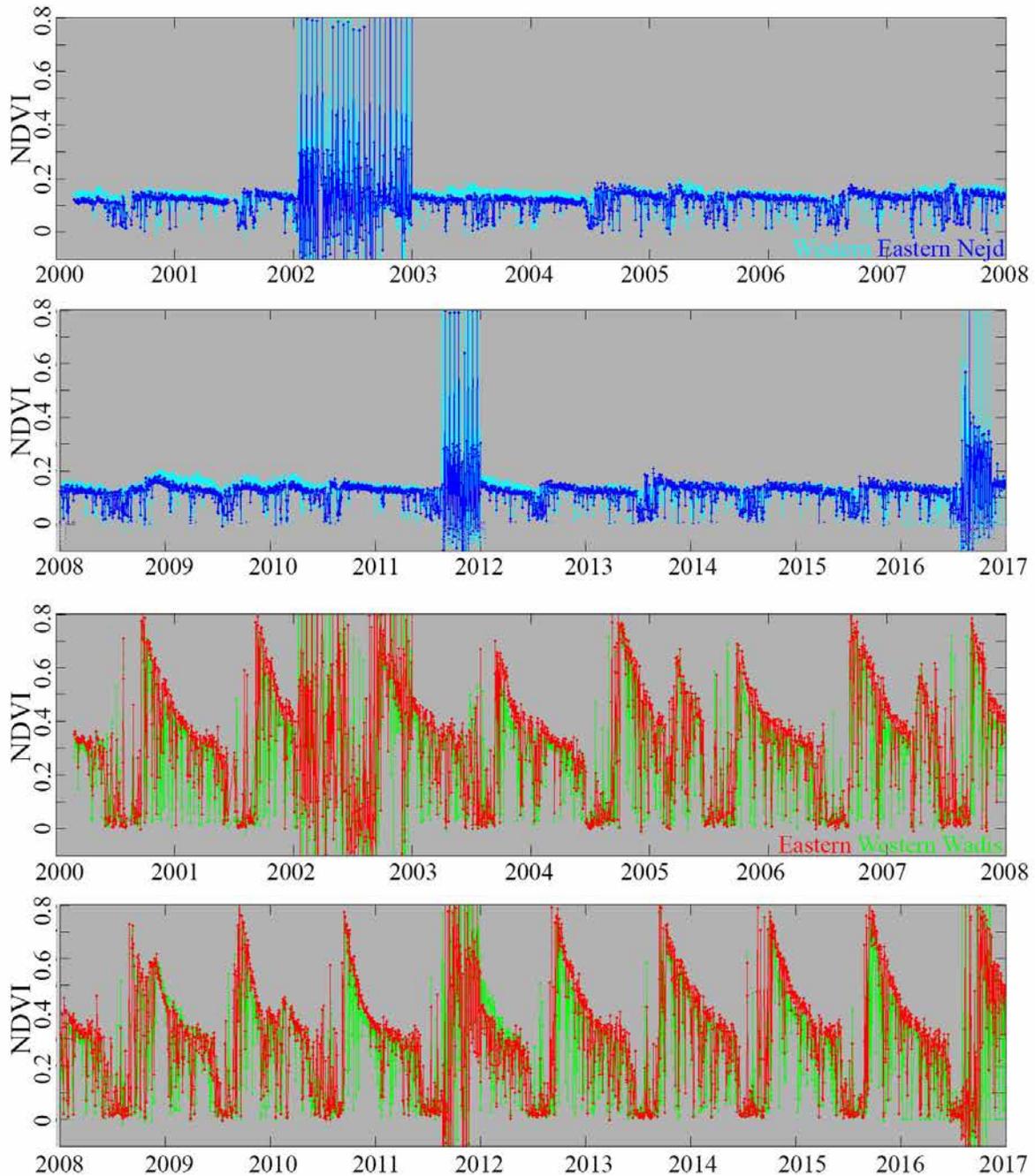

Figure 9b. Full time series for spatial averages of 101 pixel spectra from the Jabal Qamar. As in the Qara, substantial variability is observed both between EMs and within each EM across years. Eastern wadis are generally greener than western wadis. Again, wadis vary considerably from year to year. Post-khareef regreening appears more frequent and more intense in the Qamar than the Qara. As in the Qara, years with both strong and weak greening of the Qamar are present at the beginning and end of the time series. Again, no clear trend in EM time series observed in terms of duration or timing of khareef, maximum greenness, or rate of senescence of either EM phenology. Minor but consistent differences are also present in the eastern and western Nejd. The values of NDVI involved are so low that we do not attempt to ascribe physical meaning to these differences.

Year-to-year variability is shown using the full 17 year time series for spatial averages of each EM region (Figure 9). Full time series, including the years with obvious data quality issues (2002, 2011, and 2016) are shown to illustrate the character of the dataset.

As observed in the mean and variability plots shown in Figure 8, the western rangetops of the Qara (magenta) are consistently greener longer than the eastern rangetops (cyan). The central rangetops (yellow) generally lie between. Some years show prominent longitudinal differences, while for other years all 3 rangetop regions behave nearly identically. The same longitudinal pattern is even more pronounced for the eastern and western wadis.

Much of the year-to-year variability occurs coherently across EM regions. Some years are substantially greener than others in every region, and in some years the greening lasts longer than others (e.g. 2004 vs 2005). However, in other years greening occurs in a spatially variable sense and is confined to individual EMs or subsets of EMs (e.g. western rangetops are substantially greener in 2009 than 2010, but eastern and central rangetops are not).

Additionally, the early parts of 2005, 2007, and 2010 show prominent early greening (or regreening) in the wadis of both ranges. This result was unexpected and, to our knowledge, this has not been previously documented.

In contrast, time series for the eastern and western portions of the Jabal Al Qamar (Figure 9b) cloud forest show prominent differences from the eastern and western Jabal Al Qara cloud forest. One aspect of this difference is the nature of cloud cover. The duration of persistent khareef cloud cover is substantially shorter in the Qamar than the Qara. However, much more day-to-day variability is observed in the non-khareef Qamar than the Qara. While this is partially due to the smaller sample size used in the averaging (101 vs 1001 pixels), this also may be due to more frequent non-khareef cloudiness in the Qamar than the Qara. This non-khareef cloudiness in the Qamar was observed in both field seasons of the authors

Both regions in the Qamar also show generally higher amplitude phenology than the Qara. Qamar EM regions reach higher NDVI values than the Qara, greenness persists later into the dry season, and early greening is substantially more prominent in the Qamar than the Qara.

Finally, as discussed above, the two Nejd time series show little structure and are merely included for completeness.

*Analysis of Trends in Spatiotemporally Distinct Regions*

**Figure 10a. Interannual trends in vegetation cover for wadi cloud forests. The Julian day with the fewest missed (or cloudy NDVI < 0.1) observations was chosen for three 30-day post-khareef time windows. Anniversary time series for spatial averages of each region are plotted, along with best fit linear trends. Only one of the trends is significant to 95% confidence, but the slope is 0.02. The number of years in the time series is modest, some early years of the time series are likely affected by cloud contamination, and no trend is apparent in that EM either earlier or later in the decade. For these reasons, we do not ascribe any physical significance to this trend without further evidence.**

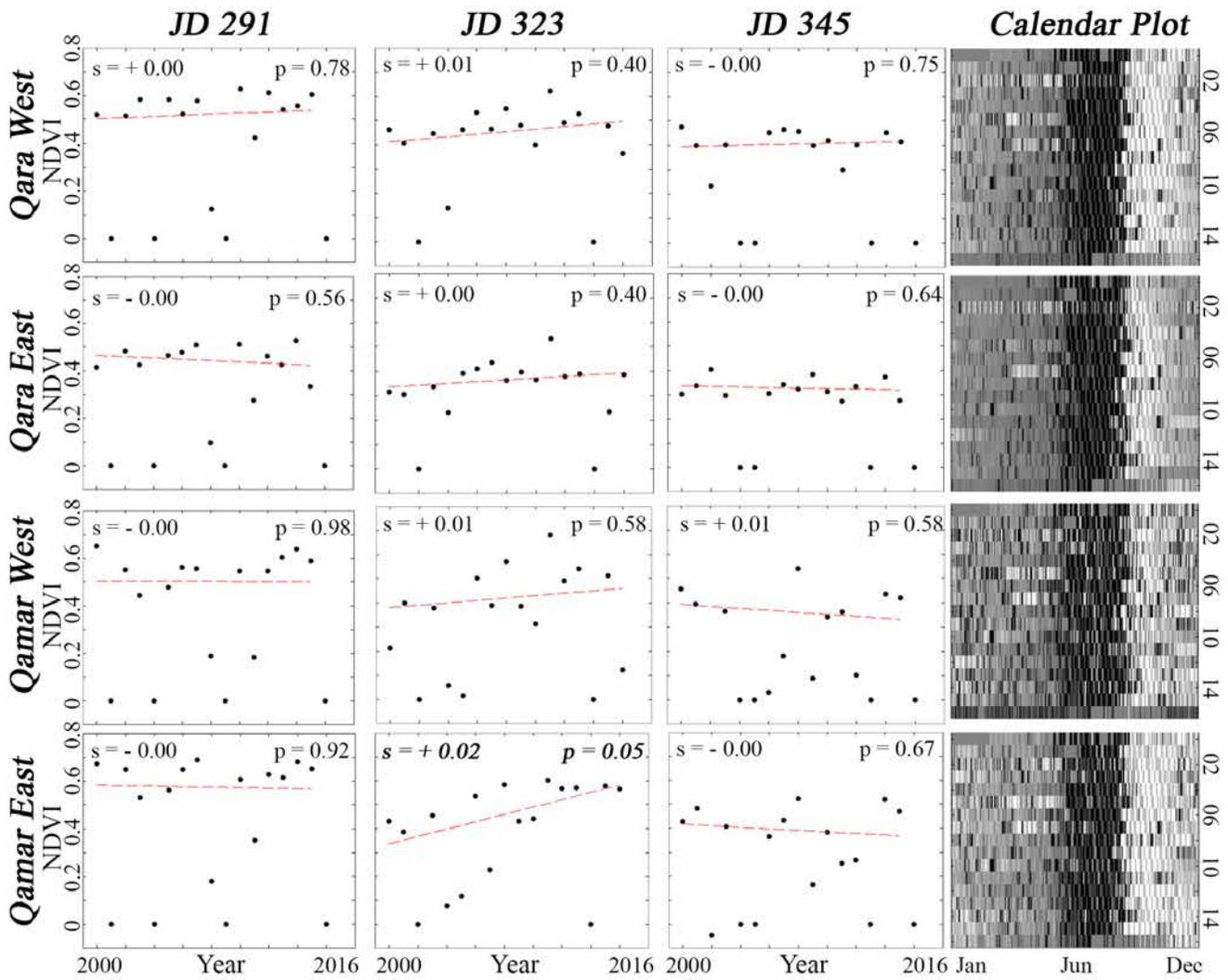

Figure 10a. Interannual trends in vegetation cover for wadi cloud forests. The Julian day with the fewest missed (or cloudy NDVI < 0.1) observations was chosen for three 30-day post-khareef time windows. Anniversary time series for spatial averages of each region are plotted,along with best fit linear trends. Only one of the trends is significant to 95% confidence, but the slope is 0.02, The number of years in the time series is modest, some early years of the time series are likely affected by cloud contamination, and no trend is apparent in that EM either earlier or later in the decade. For these reasons, we do not ascribe any physical significance to this trend without further evidence.

**Figure 10b. Same as Figure 9a, but for the pixel EM averages in the rangetops of the Jabal Qara. None of the trends are statistically significant to the 95% confidence level. Similar to the case in the wadis, interannual variability in the post-khareef phenology is substantial and dominates over any clear interannual trend.*

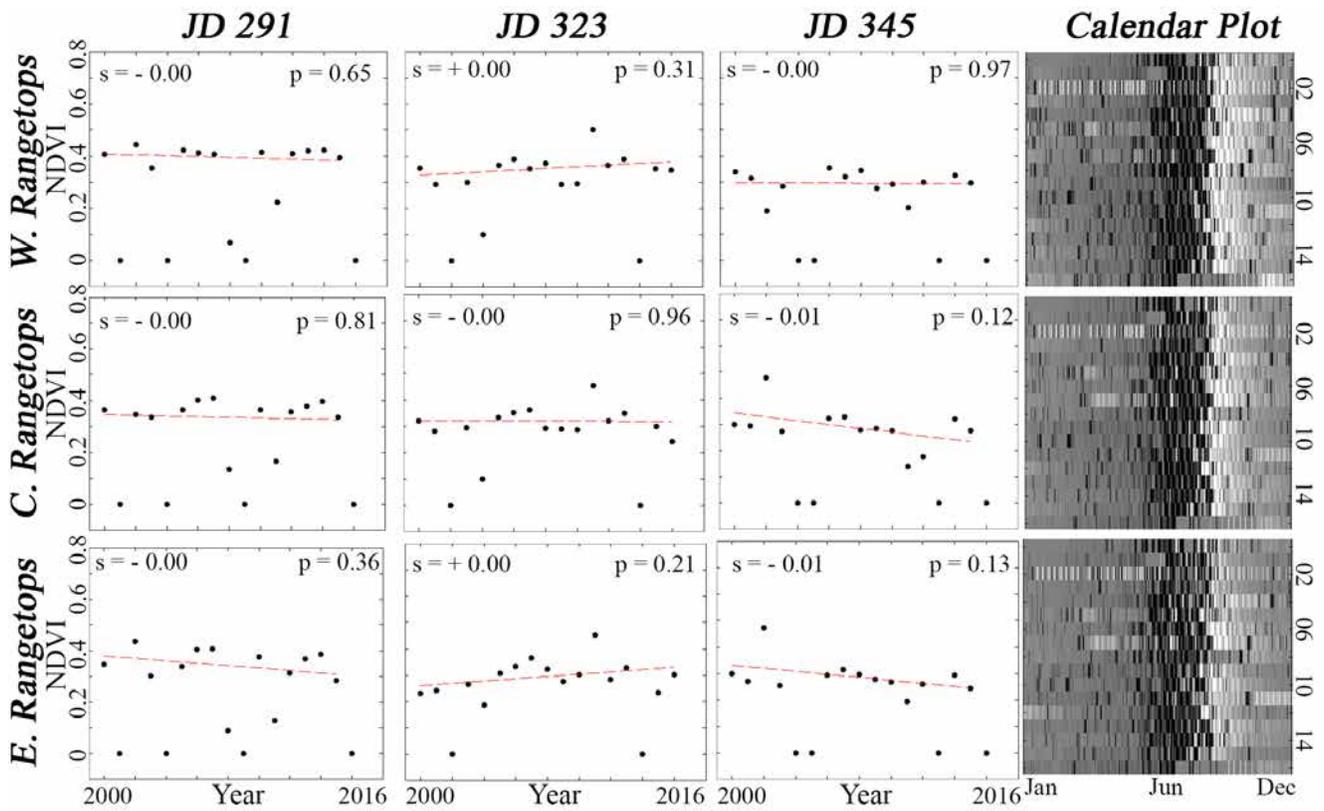

Figure 10b. Same as Figure 10a, but for the regional averages in the rangetops of the JabalAl Qara. None of the trends are statistically signifcant to the 95% confidence level. Similar to the wadis, internanual variability in the post-khareef phenology is considerable and dominates over any interannual trend that may or not be present.

We next investigate the question of interannual trends in individual EM regions. We compare EM regions, rather than a spatial average of an entire range, because the phenological distinctions discussed above clearly result in different interannual patterns (Fig. 9). For an interannual trend to be detected, its magnitude must be at least as large as the interannual variability in the phase and amplitude of the phenological cycle. We approach this question by using anniversary images (collected on the same Julian day each year) from days with the fewest number of cloudy acquisitions across the 17 year time series. We do this for three 30-day windows after the end of the khareef (Figure 10).

Anniversary images are required because of the observed natural interannual variability in timing and amplitude of the coupled monsoon-phenological system. As observed by the strength of the seasonal cycle shown in Figures 8 and 9, introducing even small changes in timing of the phenological cycle have the potential to bias multiyear statistical analysis.

We choose the months after the khareef in order to optimize the tradeoff between strength of NDVI signal and frequency of cloud cover. However, to avoid introducing bias by choosing an unrepresentative Julian Day, we also show full calendar plots for each EM region in Figure 10.

For these 17 x 366 matrices, every cell of the matrix represents one day of the time series. Every row is a year and every column is a Julian Day. Spatial average NDVI of each set of EM pixels is represented as grayscale brightness. Brighter cells have higher NDVI and darker cells have lower NDVI. These calendar plots have the advantage of displaying the entire image time series for each EM region. Obvious trends in the khareef or vegetation phenology should be apparent from visual examination of the calendar plots. Less obvious trends may emerge from regressions of the anniversary date time series. The anniversary plots shown to the left of the calendar plots are simply column profiles of the calendar plots with the fewest data drop-outs or cloud cover.

Of the 21 anniversary date – EM region combinations examined here, only 1 shows a trend significant at the 95% confidence level. This single trend implies a very small (0.02 NDVI / year) progressive increase of wadi vegetation cover in the Eastern Qamar. However, we do not ascribe any physical meaning to this trend because the meaning of 95% confidence implies that, on average, when examining 20 time series 1 will show a significant trend based on random chance. Furthermore, the small amplitude of the trend pushes the detection limit of the NDVI metric beyond our confidence.

The lack of any apparent trend in the daily MODIS NDVI directly contradicts the conclusion of a recently published study [*Galletti et al.*, 2016] which infers widescale land degradation throughout the Jabal Dhofar from an analysis of 16-day MODIS NDVI composite imagery. This disparity between daily and composite data is discussed in detail below.

*Characterization of Monsoon Cloud Cover in Spatiotemporally Distinct Regions*

**\*\*Figure 11. Interannual comparison of cumulative NDVI relative to mean for each EM region. Values decrease during the first half of the time window because frequent cloudcover during the khareef results in NDVI below the 17-year mean. At the end of the khareef, cumulative NDVI abruptly begins to increase as clouds clear to reveal dense, green vegetation. This increase continues at variable rates for variable lengths of time across regions and years. Analysis of the mean of the 17-year time series is shown in the last panel, describing the average behavior of cumulative NDVI for each region. For both wadi regions, the average date of cloud clearing is JD 257, and for all three rangetop regions the average**

*date is JD 249 (8 days earlier). In both wadis and rangetops, western regions generally have both more consistent cloud cover during the khareef and more intense and long lasting vegetation after the khareef. Years 2002, 2011, and 2016 have data quality issues but are included here for completeness.*

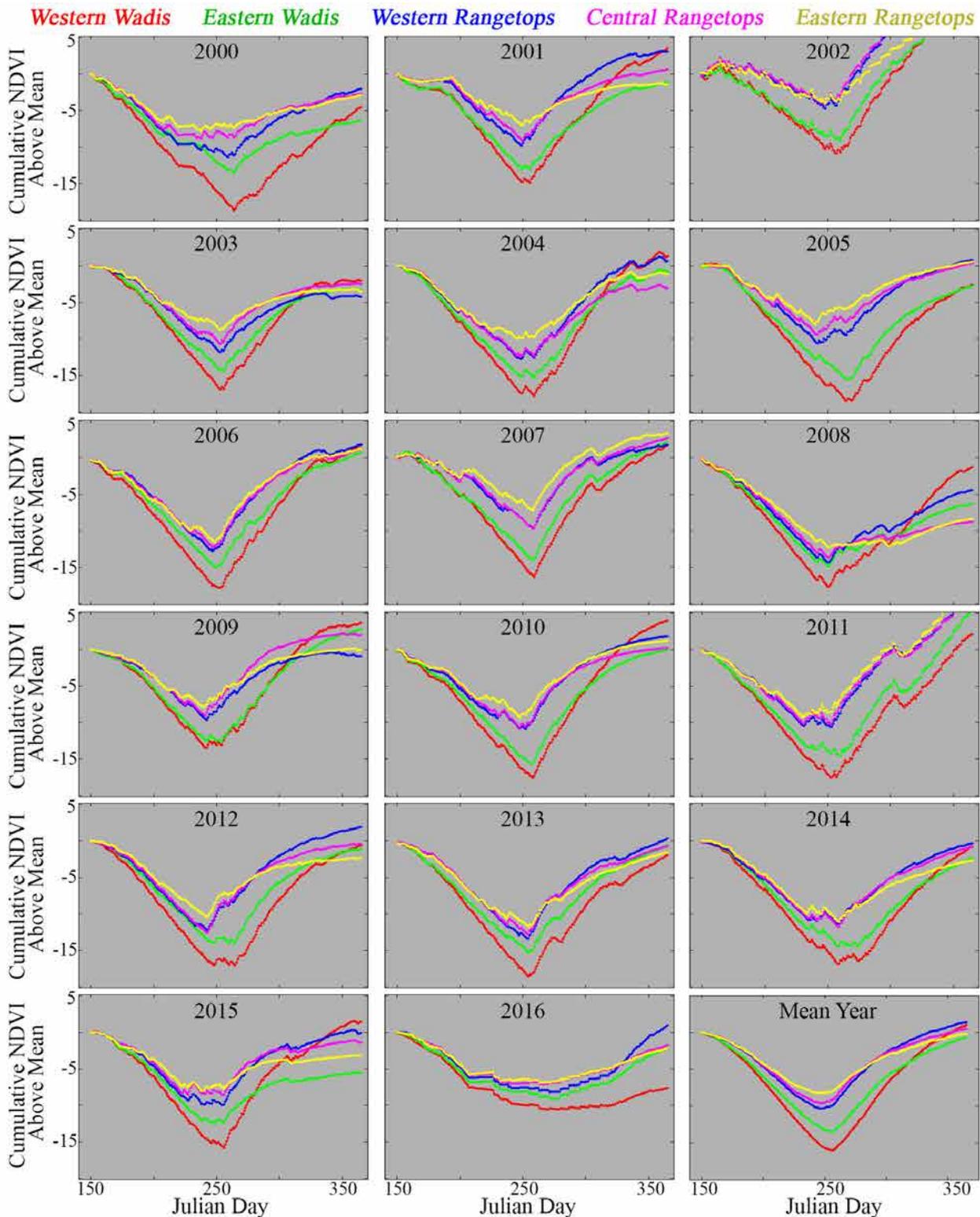

Figure 11. Interannual comparison of cumulative NDVI relative to mean for each EM region. Values decrease during the first half of the time window because frequent cloudcover during the khareef results in NDVI below the 17-year mean. At the end of the khareef, cumulative NDVI abruptly begins to increase as clouds clear to reveal dense, green vegetation. This increase continues at variable rates for variable lengths of time across regions and years. Analysis of the mean of the 17-year time series is shown in the last panel, describing the average behavior of cumulative NDVI for each region. For both wadi regions, the average date of cloud clearing is JD 257, and for all three rangetop regions the average date is JD 249 (8 days earlier). In both wadis and rangetops, western regions generally have both more consistent cloud cover during the khareef and more intense and long lasting vegetation after the khareef. Years 2002, 2011, and 2016 have data quality issues but are included here for completeness.

Finally, we use the cumulative NDVI metric explained earlier (Figure 4) to quantify the relationship between amplitude and timing of the khareef and the subsequent phenological cycle for the 5 EM regions of the Jabal Al Qara (Figure 11). The cumulative NDVI, relative to its annual mean ($\mu$), shows the relationship between cloud cover (NDVI < $\mu$) and greening (NDVI > $\mu$) as the relationship between the decreasing sum (cloud) and the increasing sum (green vegetation).

For each calendar year, the metric begins at zero then decreases to larger and larger negative values as the vegetation has already (mostly) senesced below its average value. The decrease accelerates as the persistent cloud cover of the khareef results in even lower NDVI values (< $\mu$. The cumulative NDVI metric reaches its minimum when the clouds clear to reveal a green landscape with high NDVI values. Consecutive cloud free days with abundant green vegetation cause the cumulative sum to increase rapidly. The rate of this increase decays over time as the vegetation senesces and is grazed.

We observe substantial variability from year-to-year and across EM regions in the timing, duration, and consistency of khareef cloud cover. The wadis show longer, more persistent khareef cloud cover (i.e. deeper troughs) than the rangetops. The western wadis and western rangetops show more persistent khareef than their eastern counterparts. The central rangetops vary from year to year between close similarity to the eastern rangetops, close similarity to the western rangetops, and values in between.

Some years (e.g. 2015) show a weak khareef across all EM regions, others (e.g. 2010) show a strong khareef across all EM regions, and others are in between. The mean year for each EM region is shown in the lower right. The relative amplitude of the khareef in each EM region is in strong agreement with the relative amplitude of the phenological cycle in those regions. As in Figure 9, the years with data quality issues (2002, 2011, and 2016) are included for illustrative purposes. The comparisons in Fig. 11 concisely summarize the geographic and interannual variability in the relationship between the duration and extent of khareef cloud cover and the subsequent response of the vegetation communities to the precipitation delivered during the khareef.

### *Discussion:*

The EOF + TFS spatiotemporal characterization method presented here can simultaneously quantify monsoon cloud cover and vegetation phenology objectively while requiring minimal assumptions *a priori*. In the Jabal Dhofar, applying this method on daily MODIS imagery yields spatially contiguous, temporally distinct and geographically consistent phenological patterns. The monsoonal character of this study area allows the same analysis to characterize geographical patterns and interannual changes in both climatic and ecological systems. The agreement between spatial patterns of EM affinity and known features of the physical geography of the Jabal Dhofar provides strong evidence that the spatiotemporal patterns identified from the feature space are physically meaningful.

The EOF + TFS method is sufficiently general to be broadly applicable to synoptic spatiotemporal analysis of other systems, especially other systems in which seasonal precipitation influences vegetation phenology. Interannual variability across and within EMs can be explicitly quantified for statistically distinct spatiotemporal regions, allowing for differentiation between distinct sets of temporal trajectories. Understanding of the response to the spectral index to cloud cover as well as vegetation phenology allows this variability to be simultaneously characterized for a) the onset and duration of khareef, as well as b) the amplitude & extent of vegetation phenology.

Based on this analysis of daily MODIS, as well as a forthcoming parallel analysis of Landsat-derived land cover [*Small et al.*, 2017] we find no evidence for the claims of satellite-observed degradation found by [*Galletti et al.*, 2016]. While our field observations do confirm extensive overgrazing throughout the study area, we see no evidence that this degradation is captured in the MODIS record. We do not interpret our results as contradicting claims of overgrazing, but rather state that the effects of overgrazing are apparently masked by concurrent senescence of post khareef vegetation greening and interannual variability of phenology related to variability in the timing and extent of the khareef.

Finally, the inferred link between cloud cover and vegetation phenology found in this analysis relies on soil moisture. However, confirmation of the inferred cloud cover-soil moisture-vegetation abundance causal mechanism requires field validation. One way to test this hypothesis could be continuous *in situ* measurement of soil moisture at sites controlled to have a minimum of disturbance from humans or livestock. These measurements could then be compared to ongoing near-daily observations of cloud cover and vegetation abundance as used in this study. This hypothesis predicts that soil moisture should increase with the onset of khareef cloud cover and begin to subsequently decrease when the clouds clear and evapotranspiration increases. Regions with higher soil moisture would be expected to have more frequent or longer duration cloud cover, as well as more abundant vegetation and/or slower rates of senescence. Because this investigation would only strictly require relative (and not absolute) measurement of soil moisture, relatively inexpensive frequency-domain reflectometry probes could be used.

*Conclusions:*

We use 17 years of daily MODIS NDVI imagery to characterize spatiotemporal dynamics of monsoon cloud cover and vegetation phenology in the Jabal Dhofar, southern Oman. The spectral index clearly distinguishes clouds, substrates and green vegetation, thereby quantifying the temporal evolution of the monsoon phenology at daily temporal resolution. The use of daily observations avoids the spatial and temporal aliasing inherent in the 8 and 16 day composites commonly used with MODIS vegetation indices. By maximizing both the spatial and temporal resolution, the daily MODIS product captures the rapid greening and senescence of post-monsoon vegetation and the abrupt tranisitions between rangetop grasslands and wadi cloud forests. Using the EOF spatiotemporal characterization, we find five distinct spatiotemporal phenologies in the Jabal Al Qara and two in the Jabal Al Qamar, with clear gradients in slope (between wadis and rangetops) and longitude (between east and west) in each range. The distribution of these temporal endmember patterns is spatially coherent, indicating that the structure of the temporal feature space represents meaningful geographic variations in cloud cover and vegetation phenology. We find substantial interannual variability in monsoon cloud cover and vegetation phenology in every spatiotemporal pattern and no significant trend in cloud cover or phenology in any of the regions. Finally, using the degree of asymmetry of the cumulative NDVI time series, we infer a causal link between interannual observations in duration of monsoon cloud cover and amplitude of vegetation phenology. While we cannot conclude a definitive relationship at this time, this topic presents an attractive avenue for future work.